\documentclass[12pt,preprint]{aastex}

\newcommand{\oxysix}{O~{\footnotesize{VI}}}  	
\newcommand{\oxyfive}{O\(^{+5}\)}	
\newcommand{\nitfour}{N\(^{+4}\)}	
\newcommand{\carthree}{C\(^{+3}\)}	
\newcommand{\cartwo}{C\(^{+2}\)}	
\newcommand{\hone}{H~{\footnotesize{I}}}  	
\newcommand{\hi}{H~{\footnotesize{I}}}	
\newcommand{\copernicus}{{\it{Copernicus}}}	
\newcommand{\fuse}{{\it{FUSE}}}			
\newcommand{\orfeus}{{\it{ORFEUS}}}		
\newcommand{\orfeusii}{{\it{ORFEUS\ II}}}	
\newcommand{\rosat}{{\it{ROSAT}}}		
\newcommand{\voyager}{{\it{Voyager}}}		
\slugcomment{accepted manuscript}
\shorttitle{Simulations of Supernova Remnants III}
\shortauthors{Shelton}
\begin{document}

\title{Simulations of Supernova Remnants in Diffuse Media III.\\
	The Population of Buoyant Remnants Above the Milky Way's Disk}


\author{R. L. Shelton\altaffilmark{1}}
\affil{Department of Physics and Astronomy, the University of Georgia,
	Athens, GA 30602}
\email{rls@hal.physast.uga.edu}

\begin{abstract}
We model SNRs at a variety of heights above the disk with a
detailed numerical simulation that includes non-equilibrium
ionization and recombination and follows the remnants' evolution 
until their hot bubbles have cooled.
We 
analytically calculate the bubbles'
buoyant acceleration and frictional drag.
%
%
From the simulation results, combined 
with the rates for isolated supernova explosions
above a height of 130~pc, we estimate
the time and space average \oxyfive, \nitfour, and \carthree\
column densities and emission intensities,
$1/4$~keV soft X-ray surface brightness, 
area coverage, and volume occupation due to the population of
isolated SNRs above the Galaxy's HI layer.
%
Irrespective of assumed supernova explosion energy, ambient
nonthermal pressure, or frictional drag coefficient used in
the calculations,
the predicted \oxyfive\ column density as a function
of height matches the observed distribution between 130~pc
and 2000~pc.
%
The \oxysix\ resonance line emission (1032, 1038 $\lambda\lambda$)
contributes significantly to the average observed intensity.
%
Assuming our modest supernova explosion rate,
the population of isolated extraplanar SNRs 
can explain $80\%$ of the observed $1/4$~keV surface brightness
attributed to the extraplanar gas beyond the \hone\ layer in
the southern hemisphere.
Within the range of uncertainty in
the SN rate, such SNRs
can explain all of this observed emission
(400 counts s$^{-1}$ arcmin$^{-2}$).
Thus, extraplanar SNRs could be the most important sources
of hot gas between the Local Bubble and $z \sim 2000$~pc in
the relatively quiescent southern hemisphere.
These results stand whether the remnants are assumed to be
buoyant or not.
The population of old extraplanar SNRs should cover most, but not
all of the high latitude sky, thus explaining the mottled appearance
of the soft X-ray maps (outside of superbubbles).
Bright young extraplanar SNRs should cover less than $1\%$ of the
high latitude sky.   Perhaps
the $\ell=247^{\rm{o}}, b=-64^{\rm{o}}$ crescent in the
$1/4$~keV X-ray maps could be such a remnant.
%
%
\end{abstract}


\keywords{hydrodynamics -- Galaxy:general --
ISM:general -- ISM:supernova remnants
ultraviolet:ISM -- ultraviolet:ISM}


\section{Introduction}

About half of the progenitor stars at the Sun's Galactocentric radius 
reside above the Galactic \hone\ disk
\citep{ferriere98}.  
After they explode, their remnant bubbles will evolve in a
relatively rarefied, dust-poor environment. 
They will expand to greater sizes, have lower interior pressures,
lower luminosities, lose less energy via emission by dust
grains, and live longer than disk remnants \citep{cioffi}.
At any given time, 
most of the supernova remnants (SNRs) above the \hone\ layer should be
dynamically old, having transitioned from
the adiabatic to the radiative phase in the first few
percent of their lifetimes.   Old remnants should contribute to the
\oxyfive\ and soft X-ray backgrounds even though they are dim
(\citet{shelton98,shelton99}, hereafter Papers I and II).   
The young remnants should be 
much brighter and so easier to 
recognize.  Known examples of comparatively young extraplanar supernova
remnants include SN1006 at $z \sim 450$~pc \citep{lrmb}
and the Lupus Loop at $z \sim 330$~pc 
\citep{lnh}.
In addition, bright arcs in the \rosat\ $1/4$~keV maps at 
$l = 247, b = -64$ and $l = 215, b = -68$ may also be extraplanar SNRs.

The plenitude of isolated SNRs above the Galactic \hone\ layer 
raises the possibility that
they play important roles in heating and ionizing the 
extraplanar interstellar medium (ISM) 
and whets our appetite for more information.  
Considering that SNRs evolving in low density environments 
can be 
longlived,
might they cover a large solid angle on the high latitude sky or 
fill a large fraction of the volume of space?
Might they create many of the observed high latitude high-stage ions
(\oxyfive, \nitfour, and \carthree) and soft X-rays?
Considering that young SNRs are both bright in soft X-rays and
rich in \oxyfive\ ions, do soft X-rays and 
\oxyfive\ ions track each other in general, or only in 
recently shocked gas?   How buoyant is the hot gas within SNRs?

Much progress has been made toward understanding these SNRs.
According to \citet{cioffi}'s and Paper I's rough estimates,
SNRs fill a small fraction of the extraplanar volume,
but cover a large fraction of the high latitude sky.
\footnote{This paradox is possible when a large number of SNRs 
exist at any given time and each is small on the scale of the
available space (fractional area coverage goes as 
$N (R/L)^2$, but fractional volume occupation goes 
as $N (R/L)^3$, where $N$ is the number of
objects and $R/L$ is the object's radius divided by the 
length scale of the available space, thus $R/L < 1$.)}
Paper I's preliminary estimate for the number of \oxyfive\
ions produced by the population of isolated, extraplanar SNRs
accounted for the extraplanar \oxyfive\ observed by \copernicus\ 
\citep{jenkins1978a} and the supernova explosion scaleheight 
($300$~pc)
matched the observationally determined \oxyfive\ scaleheight
\citep{jenkins1978b}.
Subsequently, \fuse\ observed a much larger sample of high
$z$ stars and found more \oxyfive\ above $300$~pc.
These new observations
suggest that modeling the vertical distribution of SN explosions
may be important.
\fuse\ also found \oxyfive\ moving away from the Galactic
plane at more than 100 km sec$^{-1}$ \citep{wakkeretal03}.
This suggests that buoyancy may be important.
Another possible sign of buoyant motion in the interstellar medium is
an \hone\ ``mushroom-shaped cloud'', whose cap may be the squashed shell of a
buoyant SNR and whose stem may consist of trailing 
material \citep{englishetal00}.
Multidimensional simulations have been able to reproduce mushroom shapes
and buoyant acceleration 
\citep{jones,avillezmaclow}.
Could buoyancy explain the large \oxyfive\ column densities and scale height?

If we could envision a map of the sky based on the
\oxyfive\ column density in extraplanar SNRs, it would not look 
like a similarly constructed map of \oxyfive\ resonance line emission,
or, for that matter, like a similarly constructed map of soft X-ray emission.
The \oxyfive\ ions would be more smoothly distributed and
cover greater area than the detectable emission.   The reason is that
SNRs harbor rich stores of high-stage ions throughout their lives,
but are bright emitters only during their early 
stages when the dense, recently shocked gas behind
their shockfronts is hot
\citep{slavin-cox-92,cioffi,shelton98,shelton99}.
Probably other types of hot gas structures behave similarly.
This difference may explain the difficulty in matching the 
\fuse\ \oxyfive\ column density map \citep{savageetal03} with
the \rosat\ $1/4$ keV map \citep{sefmpsstv}.
Furthermore, extraplanar SNRs are better emitters of 
$1/4$ keV photons than $3/4$ keV photons.   Thus, the more
smoothly distributed $3/4$ keV emission \citep{mbsk,sefmpsstv} cannot
be attributed to isolated SNRs, even though some or all of the
lumpy $1/4$~keV emission can 
(Paper II, \citet{smh}).

The Paper I and II
analyses were preliminary
in the sense that they relied on simulations of SNRs born at
a single height above the plane 
and were meant to be followed by a suite of SNR simulations.
The \citet{smh} project used the entire SNR population rather
than the population of isolated SNRs above the Milky Way's
thick disk.
Furthermore, their simulations did not include ambient magnetic pressure
which can compress old SNRs and therefore increase their
UV and X-ray luminosities.   Consequently, these two
studies predicted significantly different time and space average
$1/4$~keV surface brightnesses.   We resolve these difficulties
in the present paper.

Here, we aim to estimate the observable effects of the 
population of extraplanar SNRs by
simulating a suite of remnants at various
heights above the Galactic plane and various magnetic field
strengths and explosion energies, allowing for buoyancy, and
by multiplying by the expected rate of isolated supernova explosions above the 
Galactic disk.
The simulation package used 
to calculate the time evolution of isolated supernova remnants 
is described in detail in Paper I and described briefly in Subsection 2.1.
The analytic calculations used to predict the buoyant acceleration 
are spelled out in detail in Subsection 2.2.
The ambient conditions and progenitor statistics 
are presented in Subsections 2.3 and 2.4.
Due to the interstellar medium's variation with height above
the plane and due to the uncertainty in the explosion energy
and ambient non-thermal pressure, we ran a suite of simulations.
Their parameters are presented in Subsection 2.5.
The resulting area coverage, volume occupation, 
\oxyfive, \nitfour, \carthree, 
and soft X-ray predictions for each of the simulated SNRs are
tabulated in Subsection 3.1.
We combined the individual SNR predictions with the 
buoyancy calculations and convolved with the rate for 
isolated supernova explosions above the Galactic disk's \hone\ layer.
The resulting estimates of the time and space averaged
area coverage, volume occupation, high stage ion content and emission,
and $1/4$~keV X-ray intensity due to the population of extraplanar SNRs
are presented and discussed in Subsections 3.2 - 3.5.
In Section 4, we compare the estimates with \oxyfive\ and
$1/4$~keV X-ray observations.   \oxyfive\ and soft X-ray
emission are far less affected by photoionization than
are \carthree\ and \nitfour, so are better tests on the model.
The column densities from the calculations
agree well with the \oxyfive\ column densities observed
within the first 2000~pc of the Galactic plane.  
The simulations also explain much of the observed
extraplanar $1/4$~keV X-ray emission outside of superbubbles and
explain its mottled appearance.
We summarize the results in Section 5.

\section{Simulation Method}

\subsection{Computer Simulation Package}

	The supernova simulations are performed with a 
Lagrangian mesh hydrocode whose algorithms model shock dynamics,  
nonequilibrium ionization and recombination, 
and nonthermal pressure.	
Previous investigations
found that some form of mixing is necessary in order for 
the remnants' centers to retain the modest gas densities surmised from
X-ray observations of post Sedov phase SNRs such as W44.
The hydrocode models thermal conduction, which diffuses
entropy within the remnant's interior and so approximates the required 
mixing (\citet{cox-etal-99, shelton-etal-99, shelton-kuntz-petre}).
\citet{cui-cox} found that the electrons and ions 
behind the SNR shock zone come into equilibrium early in the
remnant's evolution, long before radiative cooling becomes important.
As a result, the SNR's late term evolution (which plays the largest
role in calculating the remnant's \oxyfive\ content, area
coverage, and volume occupation) is not affected by our
approximation of instantaneous electrons and ion equilibration.
Cosmic rays affect SNRs by contributing to the pressure behind the
shock, while SNRs affect cosmic rays by accelerating them.
Cosmic rays affect SNRs by contributing to the pressure behind the
shock, while SNRs affect cosmic rays by accelerating them.
In a careful analysis of supernova remnant evolution in the
presence of cosmic rays and magnetic fields, \citet{ferriere-zweibel}
explored several detailed interactions, but, nonetheless, found 
that their simulated SNRs with cosmic rays were qualitatively
similar to their simulated SNRs without cosmic rays.
In our simulation package, we include an isotropic cosmic
ray pressure (see Section 2.5), but neglect detailed interactions.
Furthermore, we assume that 
the interstellar medium's gas phase metal abundances are approximately
solar and so adopt the abundances of \citet{grevesse_anders}.
The code is similar to \citet{slavin-cox-92,slavin-cox-93}'s code.
Papers I and II used this code,
describe it in detail, and present sample simulations.
Because the hydrocode's Lagrangian mesh architecture does not model
a stratified ambient media,
we combine the SNR simulations with the following 
analytic buoyancy calculations in order to determine if and how 
rapidly the remnants
rise in the Galaxy's stratified interstellar medium.

\subsection{Buoyancy}

	As a whole, a SNR is not buoyant, because
its average density approximately equals that of the surrounding gas.
Only the very hot, rarefied interior should be buoyant.
Furthermore, buoyancy is probably not relevant for very young remnants
because 
the buoyant acceleration acting on the bubble interiors
is outpaced by the remnants' expansion and because
their magnetic fields tie the most buoyant material in their 
centers to the non-buoyant dense, ionized, postshock gas.
Later in a remnant's evolution, a cool shell develops between
its hot bubble interior and its shockfront (see \citet{slavin-cox-92},
or Paper I for detailed simulations).
At this time, buoyant motion may exceed the hot bubble's expansion
and the postshock shell cools, becomes somewhat neutral, 
and dissipates.
It is in this stage that we consider the buoyant force.  
Because we do not know how well the shell constrains the hot interior
nor the strength of frictional drag,
we consider a range of cases.  In the first,
we assume that the SNRs do not rise buoyantly.
In the second, 
we assume that after the cool shell develops, the hot remnant interiors
dissociate from the shell and rise buoyantly, but encounter some
resistance.   
In the third, we assume that after the cool shell develops, the
hot remnant interiors dissociate from the shell and rise buoyantly,
without resistance.
Below, we describe our analytical calculations for the buoyant
motion of a hot bubble interior encountering frictional drag.

The elementary equation for the acceleration due to 
buoyancy is:

\begin{equation}
a(z) = g(z)(1 - \frac{\rho_a(z)}{\rho_b(z)}),
\label{eq:buoyancy}
\end{equation}
where $g$ is the acceleration of gravity, 
$\rho_a$ is the mass density of the ambient medium, 
$\rho_b$ is the mass density within the bubble,
and $z$ is the height above the Galactic plane.
The acceleration of gravity at the Sun's galactocentric
radius is given by
\citet{benjamin_danly} and  \citet{wolfire_etal} as:
\begin{equation}
g(z) = - 9.5 \times 10^{-9} \ \ \tanh (z/400{\rm{pc}})\ {\rm{cm\ s}}^{-2} .
\label{eq:gravity}
\end{equation}

As the bubble gains speed, it should experience increasing frictional
drag.
Following \citet{benjamin_danly}, we calculate the frictional
acceleration in the direction opposite to motion from
\begin{equation}
a_{f} = \frac{1}{2} C A_{b} v_{b}^2 \rho_{a}/m_b,
\end{equation}
where 
$A_{b}$ is the cross sectional area of the hot bubble, 
$v_{b}$ is the hot bubble's velocity relative to the ambient medium 
$\rho_{a}$ is the ambient mass density,
$m_b$ is the mass of the hot bubble (determined from following
simulations)
and
$C$ is the drag coefficient.
If the bubble slides frictionlessly through the ISM,
then $C = 0$.   However,
if the bubble sweeps up the overlying ISM, then $C = 2$. 
If, in addition, a low pressure region develops behind the bubble,
then $C$ exceeds 2.
The value of
$C$ for hot bubbles in the ISM is poorly known, but
simulations suggest \citep{jonesetal} that it may be
around 1.   Here we consider 3 cases, $C = 0$, $C = 1$, and 
$C =$ large enough to prevent bubble motion.

Generally, the center of the hot bubble is the most rarefied
and thus most buoyant, while the periphery of the hot bubble
is the densest and least buoyant.
A strong buoyancy gradient develops which may allow the
interior to rise relative to the periphery.
In our calculations of the mass and area of the hot bubble, 
we include only the buoyant region of the hot bubble.
For computational simplicity, we assume that the buoyant region
remains intact.   Thus, we calculate the force on this region
and divide by its mass to find its average acceleration.
We then iterate using small time steps to find the velocity
and vertical displacement.   

\begin{figure}           
  \plottwo{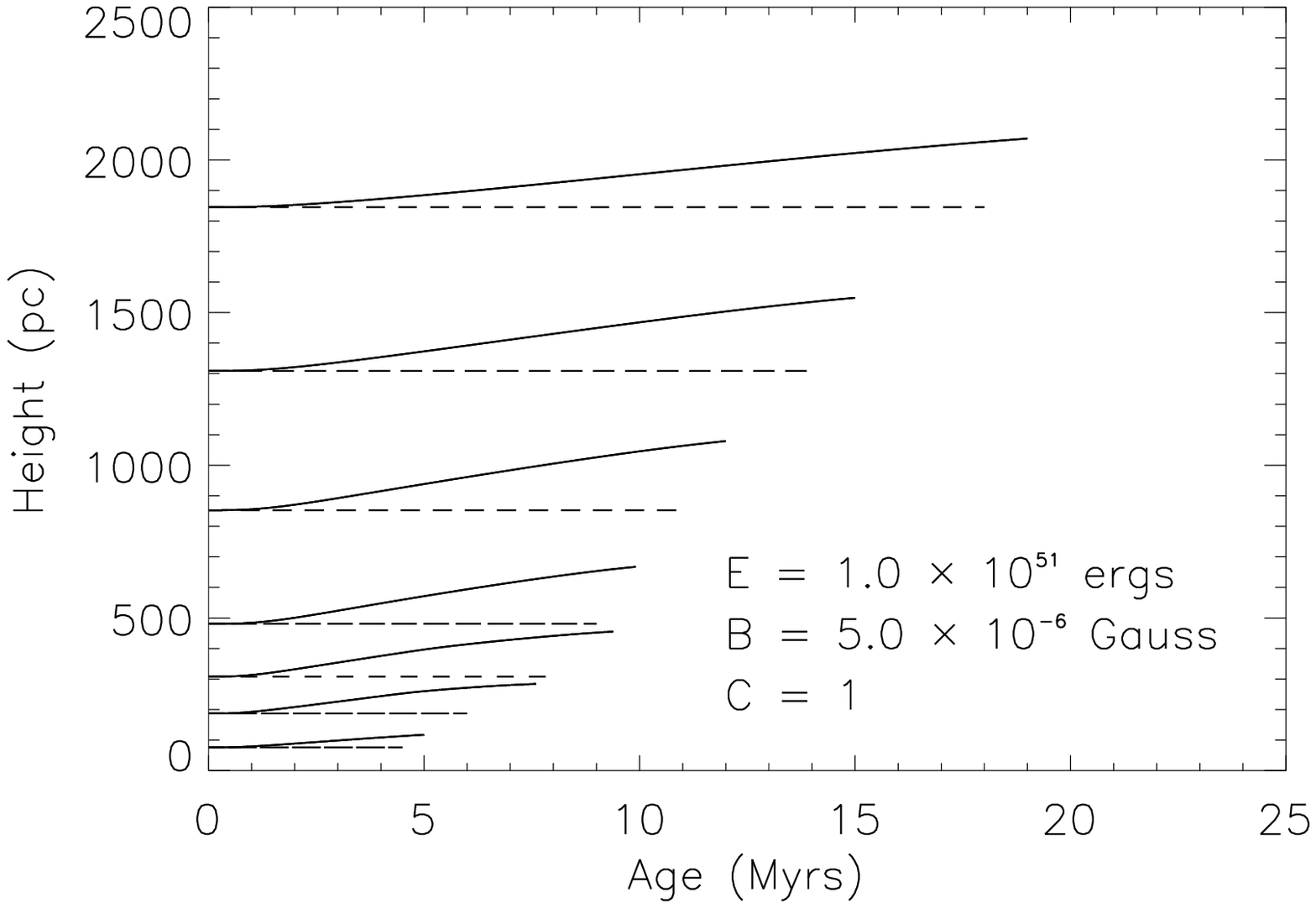}{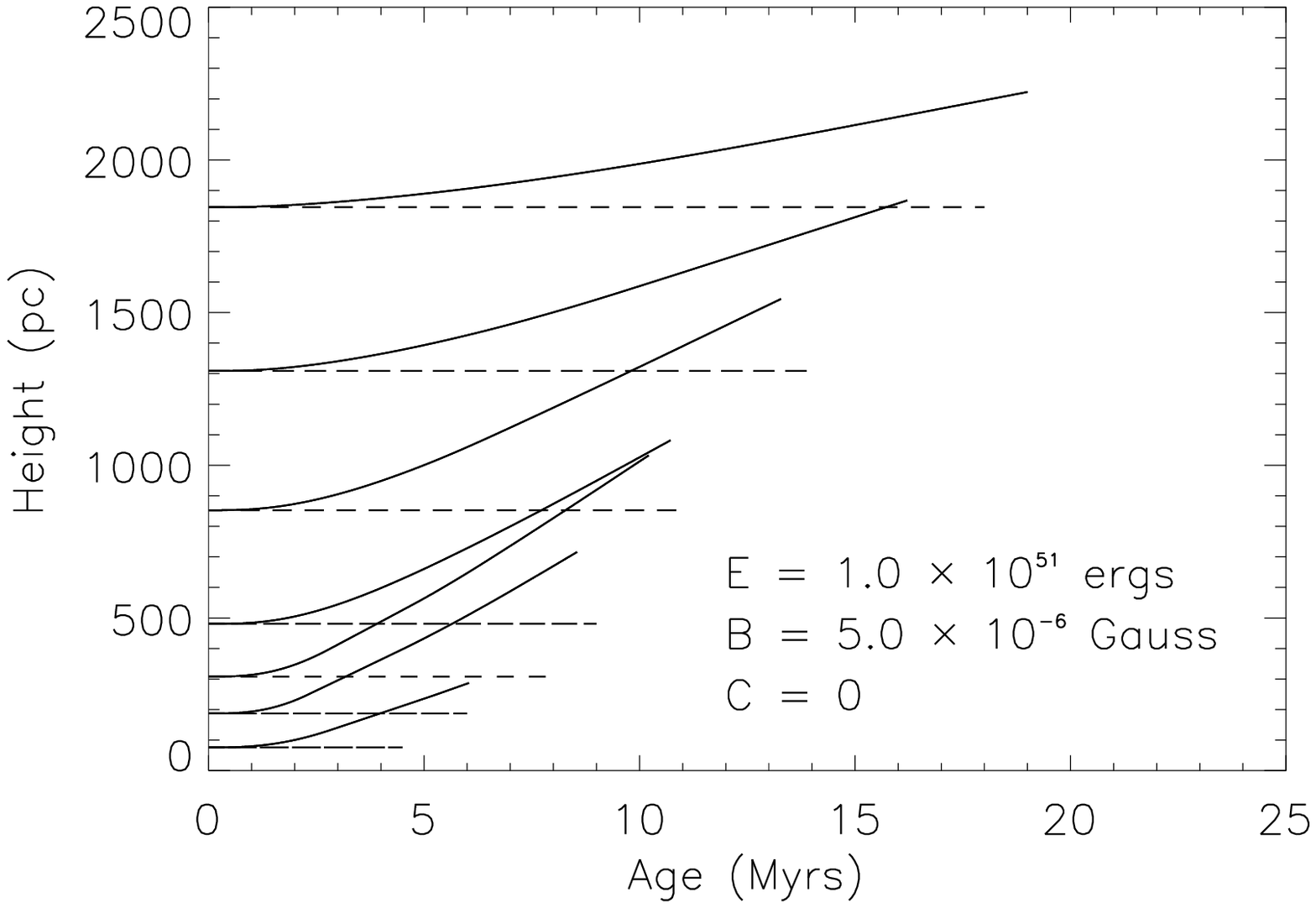}
  \caption[Buoyant Rise]
{In each figure, the dashed lines track the non-buoyant SNRs
and the solid lines track the buoyant SNRs as they rise.
In the left and right figures, respectively, the drag coefficient, $C$, 
which is used to predict the trajectories of the buoyant SNRs,
is set to 1.0 and 0.0, respectively.
The height refers to the height of the center of the remnant
with respect to the Galactic plane.
The height gains for this choice of simulation parameters
(explosion energy $= 1.0 \times 10^{51}$~ergs, effective 
magnetic field strength $= 5.0\ \mu$G)
lie between those for the other two sets of simulation parameters 
used in this paper and described in Section 2.5.}
\label{fig:buoyancy}
\end{figure}

As the bubble rises, it passes through successively more
rarefied ambient gas.   
In some cases, a bubble will rise from its birth neighborhood
into the overlaying neighborhood.   At that point we account 
for the new environment's effects on the remnant by
switching to the simulation results for a remnant
evolving in the new environment.  Thus, we use the first
part of the simulation for a remnant in the birth neighborhood
and the second part of the simulation for a remnant in the
overlaying neighborhood.
We must deal with the possibility that the two remnants may have
differing lifetimes and hence differing maturation rates.
Although the aging process may not be strictly linear, 
we define a remnant's ``maturity'' as the ratio of the
present age to the total lifetime, where
the remnant's lifetime is the time required for all of the material
in a stationary remnant to cool below $2 \times 10^4$~K (see Paper I
for further discussions of remnant death).
When we switch from the remnant evolving in the lower neighborhood to 
the remnant evolving
in the overlaying neighborhood, we match maturities rather than ages.
Later in this article, we will perform time 
integrations.   For the integrations, we will not count the
time gap between the age of the first remnant and the age
of the second.   If we were to count that time, we would
overestimate the effects of the remnants.
Nonetheless, when we allow the remnants to be buoyant, they
live longer than stationary remnants, as shown in 
Figure~\ref{fig:buoyancy}.

\clearpage

\clearpage

\subsection{The Ambient Medium in the Thick Disk and Lower Halo}

Both the hydrocode simulations and the buoyancy calculations require
estimates of the ambient density of material.  From \citet{ferriere98},
the vertical distributions of molecular, neutral, warm ionized, and hot hydrogen 
nuclei are:
\begin{equation}
n_m(z) = 0.58 \ e^{-(z/{81 {\rm{pc}}})^2} \ {\rm{cm}}^{-3}
\label{eq:n_mol}
\end{equation}
\begin{equation}
n_{n}(z) = (0.395 \ e^{-(z/127{\rm{pc}})^2} + 0.107 \ e^{-(z/318{\rm{pc}})^2}
+ 0.064 \ e^{-|z|/403{\rm{pc}}}) \ {\rm{cm}}^{-3}
\label{eq:n_n}
\end{equation}
\begin{equation}
n_{wi}(z) = (0.0237 \ e^{{-|z|}/{\rm{1\ kpc}}} 
+ 0.0013 \ e^{{{-|z|}/{\rm{150\ pc}}}}) \ {\rm{cm}}^{-3} 
\label{eq:n_wi}
\end{equation}
\begin{equation}
n_h(z) = 4.8 \times 10^{-4} \ e^{{-|z|}/{1.5 {\rm{kpc}}}} \ {\rm{cm}}^{-3}.
\label{eq:n_h}
\end{equation}
%
The ratio of helium to hydrogen atoms is approximately 1 to 10, so
the total density of atoms is assumed to be approximately 1.1 times the 
density of hydrogen atoms.  
From Equations~\ref{eq:n_mol} through \ref{eq:n_h} 
we see that 
the \hone\ scaleheight is approximately 130~pc,
half of the molecular, atomic, and ionized 
material lies below $z \sim 90$~pc,
and $80\%$ of the material 
lies below the scaleheight for isolated supernovae 
($\sim300$~pc, presented below).

\subsection{The Distribution of Isolated Supernova Progenitors}

From \citet{ferriere98}, the volumetric rate of Type 1a supernova explosions near
the Sun's Galactocentric radius is:
\begin{equation}
{\rm{Rate}_{Ia}}(z) = 4.0 \times 10^{-6} e^{-|z|/325 {\rm{pc}}}\ {\rm{kpc^{-3}yr^{-1}}.}
\label{eq:rate1a}
\end{equation}
The expected rate of isolated Type Ib plus Type II supernova explosions is
bracketed by estimates taken from \citet{ferriere98} (lower range) and 
calculated from
\citet{mw} (upper range):
\begin{equation}
{\rm{Rate_{Ib + II}}}(z) = 1.4 {\rm{\ to\ }} 2.8 
\times 10^{-5} e^{-|z|/266 {\rm{pc}}}
\ {\rm{kpc^{-3}yr^{-1}}.}
\label{eq:rate1biinew}
\end{equation}
Note that isolated explosions account for only $40\%$ of the
TypeIb and Type II rate.   The remaining explosions occur in clusters
near the Galactic midplane.   They create superbubbles,
which are well analyzed in \citet{ferriere98}, but outside our
frame of interest.
For the following calculations, we will use the total isolated SN
explosion rate at the Sun's Galactocentric radius, in 
which the isolated Type Ib plus Type II rate is taken as the
average of the \citet{ferriere98} and \citet{mw} estimates:
\begin{equation}
{\rm{Rate_{SN}}}(z) = ( 4.0 \times 10^{-6} e^{-|z|/325 {\rm{pc}}} \  
+ 2.1 \times 10^{-5} e^{-|z|/266 {\rm{pc}}} ) \ {\rm{kpc^{-3}yr^{-1}}}.
\label{eq:snrate}
\end{equation}
In more conceptual terms, every million years 
2 to 3 isolated progenitor stars explode within
an open-ended column beginning at $z = 130$~pc, extending to
$z = \infty$, having a cross sectional area of 1 kpc$^2$, and
residing at the Sun's Galactocentric radius.

\subsection{The SNR Simulation Parameters}

A remnant born at the SN progenitor
scale height will be surrounded by a medium of density, $n_{\rm{o}}$, equal to
0.1 atoms cm$^{-3}$.   Remnants born at half the scale height
will be surrounded by $\sim 0.3$ atoms cm$^{-3}$ gas while
remnants born at twice the scale height will encounter
$\sim~0.03$ atoms cm$^{-3}$ gas.
In order to account for the variation in ambient density above the
Galactic disk,  we
performed simulations for the following sample of seven ambient densities:
0.5 ~atoms~cm$^{-3}$ 
(corresponding to a height of $z_1 = 76$~pc), 
0.2  
(corresponding to a height of $z_2 = 190$~pc), 
0.1 (corresponding to a height of $z_3 = 310$~pc), 
0.05 (corresponding to a height of $z_4 = 480$~pc), 
0.02 (corresponding to a height of $z_5 = 850$~pc),
0.01 (corresponding to a height of $z_6 = 1300$~pc),
and 0.005~atoms~cm$^{-3}$ (corresponding to a height of $z_7 = 1800$~pc).
$z_7$ is roughly 6 times the scaleheight for isolated supernova
explosions.
The $n = 0.5$~atoms cm$^{-3}$ simulations were performed for
comparison with the others; they will not be included
in the extra-planar population.
Due to uncertainties in the values of the typical 
explosion energy, we performed simulations for both 
$E_o = 0.5 \times 10^{51}$ ergs and $E_o = 1.0 \times 10^{51}$ ergs
which represent the typical range of estimated explosion energies.  
At any given height, we have approximated the SNRs' environment 
as being homogeneous with a temperature, $T$, of  
$1.0 \times 10^4$~K.  At this temperature, the hydrogen in this gas is ionized, as  
is typical of the Reynolds layer and 
of the plasma surrounding and pre-ionized by supernova remnants.
The turbulent pressure is thought to be small and 
is not explicitly included in the hydrocode simulations.  
The magnetic and cosmic ray pressures 
are swept into a single term called the nonthermal pressure, $P_{nt}$.  
Observational and theoretical estimates 
of their values in the thick disk are sparse, with the
values chosen for the simulations, $P_{nt} = 1800$ and 7200~K~cm$^{-3}$,
being well within the probable range.
If $B_{eff}$ is the
effective magnetic field strength, where $P_{nt} = B_{eff}^2/(8 \pi)$,
then the chosen nonthermal pressures correlate to
$B_{eff} = 2.5$ and $5.0 \mu$G.  
We set $B_{eff} = 2.5~\mu$G only when also setting
$E_o = 0.5 \times 10^{51}$~ergs.   Thus for every choice
of ambient density in our sample, we have performed 
three simulations, the first having 
$E_o = 0.5 \times 10^{51}$~ergs and 
$B_{eff} = 2.5 \mu$G, the second having
$E_o = 0.5 \times 10^{51}$~ergs and 
$B_{eff} = 5.0 \mu$G, and the third having
$E_o = 1.0 \times 10^{51}$~ergs and 
$B_{eff} = 5.0 \mu$G.
In all, we report on 21 simulations.  Their parameters are listed in
Table~\ref{table:param}.   

\clearpage

\begin{deluxetable}{lclll}
\tablewidth{0pt}
\tablecaption{Parameters for SNR Simulations}
\tablehead{
\colhead{$z$}     
& \colhead{$n_{\rm{o}}$}
& \colhead{$E_o$} 
& \colhead{$B_{eff}$}
& \colhead{Name}\\
\colhead{(pc)}     
& \colhead{(atoms cm$^{-3}$)}
& \colhead{($10^{51}$ ergs)} 
& \colhead{($\mu$G)}
& \colhead{}   }
\startdata
76    &  0.5  & 0.5   & 2.5  & halo 65\\

''    &  ''  & ''    & 5.0  & halo 62\\

''    &  ''  & 1.0   & ''  & halo 64\\

190   &  0.2 & 0.5   & 2.5  & halo 59\\

''    &  ''  & ''    & 5.0  & halo 57\\

''    &  ''  & 1.0   & ''  & halo 58\\

310   &  0.1 & 0.5   & 2.5  & halo 92\\

''    &  ''  & ''    & 5.0  & halo 91\\

''    &  ''  & 1.0   & ''  & halo 90\\

480   & 0.05 & 0.5   & 2.5  & halo 110\\

''    &  ''  & ''    & 5.0  & halo 111\\

''    &  ''  & 1.0   & ''  & halo 109\\

850   & 0.02 & 0.5   & 2.5  & halo 100\\

''    &  ''  & ''    & 5.0  & halo 101\\

''    &  ''  & 1.0   & ''  & halo 99\\

1300  & 0.01 & 0.5   & 2.5  & halo 103 \\  

''    &  ''  & ''    & 5.0  & halo 104 \\ 

''    &  ''  & 1.0   & ''  & halo 102 \\ 

1800  & 0.005& 0.5   & 2.5  & halo 106\\

''    &  ''  & ''    & 5.0  & halo 107\\

''    &  ''  & 1.0   & ''  & halo 105\\ \hline
\enddata
\label{table:param}
\end{deluxetable}

\clearpage

\section{Simulation Results}

\subsection{Individual Isolated Remnants}

	Each computer simulation yielded detailed,
comprehensive predictions of the remnant's physical state as a 
function of time.  
The temperature, pressure, velocity, 
high ion content, high ion absorption line profile,
X-ray spectrum, and X-ray luminosity of a sample remnant 
as a function of time are presented 
in Papers I and II.  In the present paper, we are focusing on the bulk
physical and observable effects of the population of isolated extraplanar 
SNRs and so we will extract a complementary set of
information from each simulation.

First, we turn our attention to the cross sectional area covered by 
supernova remnants.
When we look up into the Earth's atmosphere we see that
some fraction of the sky is covered by the population of clouds.   
Similarly, we can image that
some fraction of the Galactic high latitude sky is covered by
the population of supernova remnants.   Determining what this 
fraction is will
help us to interpret high latitude observations of hot gas.
Theoretically, the fraction of sky covered by SNRs above a height 
{\small{$\mathcal{Z}$}} is 
$\int_{\mathcal{Z}}^\infty {\rm{Rate}}_{{\rm{SN}}}(z) 
\times (\int \pi R^2(z,t) \ dt) \ dz$, where
${\rm{Rate}}_{\rm{SN}}$ is given in Equation 10 and $R(z,t)$ is the
radius of a remnant of age $t$ and evolving at height $z$. 
In this subsection, we determine the second factor,
$\int \pi R^2 (z,t) \ dt$, the time
integrated area of a SNR evolving at height $z$. (We will use the
integral to calculate the fractional area coverage in the following
subsection.)
We approximate the integral 
as $\sum{\pi R^2(z,t)} \times {\Delta t}$.
We use up to 26 time slices for each summation and, unless stated
otherwise, we sum through the end of the remnant's life.
The radius is determined from the location of the shock front
when the remnant is young and from Paper I's bubble boundary
criterion when the remnant is old enough to have a cool shell.
According to this criterion,
previously shocked material that has cooled to a temperature less than
$2 \times 10^4$~K is considered to be part of the shell or environmental
ISM while hotter material is considered to be part of the hot bubble.
The time integrated area of each simulated SNR is presented in
Table~\ref{table:sizepersnr}.
We added a column listing the integrated area for the 
remnants' early stage, the pre-shell formation (PSF) phase, which
ends when the gas immediately behind the shockfront is no longer
heated to more than $3 \times 10^4$~K.
This column is identified by the ``psf'' notation 
in the summation.
The pre-shell formation phase is worth attention because
during this stage
the dense gas behind the shockfront is very hot and hence luminous 
in soft X-rays and UV photons.
As a result, the remnant could probably
be observed and identified as a distinct entity during this stage.
At the end of this stage, the gas behind the shockfront
undergoes rapid radiative cooling and so evolves
into a cool shell which separates the ever-expanding
shockfront from the hot, rarefied interior. 
Although the remaining hot interior (the 'hot bubble') 
can be sufficiently
hot and ionized to produce soft X-rays and UV photons,
its density and therefore its intensity will be less than previously.
As a result, an old remnant might 
not be observed or identified as a distinct entity.
Nonetheless, as we will show later, 
the vast majority of the ions (in a time integrated calculation)
result from the post-shell formation phase.
Note that ``halo 106'' has a larger pre-shell formation area integral
than ``halo 107'' and ``halo 108'' because its shell formed later.

Let us turn our attention to the volume occupied by SNRs.
At any given moment, some regions of the Galaxy 
are occupied by SNRs while others
are not.  Ignoring overlap, 
the average fraction of space at a height of $z$ which
is occupied by SNRs (the ``filling fraction'') can be estimated from:
${\rm{Rate}}_{\rm{SN}}(z) \times (\int \frac{4}{3} \pi R^3 (z,t) dt)$.
In this subsection, we determine the second factor in the overall 
integration, $\int \frac{4}{3} \pi R^3 (z,t) \ dt$, 
the time integrated volume.  We approximate this integral
(which is called the ``four-volume'' in \citet{slavin-cox-93})
as $\sum \frac{4}{3} \pi R^3 (z,t) \ \Delta t$.
Table~\ref{table:sizepersnr} presents
the time integrated volumes of individual simulated remnants.
These values will be used in the following subsection to calculate the 
filling fraction of SNR gas.

In preparation for calculating the average \oxyfive, \nitfour,
and \carthree\ column densities in the following subsection, 
we calculate
the time integrated number of ions per SNR here.
The integrals are $\int \int_0^R n_{\rm{ion}}(r,t) \ 4 \pi r^2 \ dr \ dt$,
where $n_{\rm{ion}}(r,t)$ is the volume density of the chosen ion
at the radius $r$ from the remnant's center and at a time $t$.
The integrals are approximated by summations with respect to
time and radius and presented in
Table~\ref{table:highionspersnr}.   
Note that \carthree, and to a lesser extent, 
\nitfour\ can be produced via photoionization or collisional ionization.   
Here, we report on only the collisionally ionized \carthree\ and
\nitfour.
As in Table~\ref{table:sizepersnr},
additional columns are provided for the integrals through
the pre-shell-formation (``psf'') stage of evolution.

Similarly, we calculate the energy released in \oxyfive, \nitfour,
and \carthree\ resonance line doublet photons
(\oxyfive: 1032, 1038 \AA; \nitfour: 1239, 1243 \AA;
\carthree: 1548, 1551 \AA).
The energy released by a single SNR in a particular doublet 
is approximated as the doublet's 
luminosity summed over time, $\sum L \ \Delta t$.
See Table~\ref{table:highionsemisspersnr}.
We also calculate the energy released in $1/4$ keV soft X-ray photons.
The time integrated luminosity is approximated as 
$\sum L_{1/4\ keV} \ \Delta t$.
The soft X-ray photons are scattered across a spectrum.   So, in order
to compare our predictions with observations and other calculations,
we convolve the spectra with the \rosat\ response functions for
the R1 and R2 bands and report the results in units of
\rosat\ R1 $+$ R2 counts cm$^{-2}$.   
See Table~\ref{table:xrayspersnr}. 
As in the other tables, a column is provided for the 
contribution from the pre-shell formation phase, when the
remnants are young, bright, and most easily identified as
individual objects.

\clearpage

\begin{deluxetable}{ccccc}
\tablewidth{0pt}
\tablecaption{}
\tablehead{
\colhead{Simulated}     
& \colhead{$\sum_0^{psf}{\pi R^2} \times {\Delta t}$}
& \colhead{$\sum{\pi R^2} \times {\Delta t}$}
& \colhead{$\sum_0^{psf}{\frac{4}{3}\pi R^3} \times {\Delta t}$} 
& \colhead{$\sum{\frac{4}{3}\pi R^3} \times {\Delta t}$}   \\
\colhead{Remnant}     
& \colhead{(kpc$^2$ yr)}
& \colhead{(kpc$^2$ yr)} 
& \colhead{(kpc$^3$ yr)}
& \colhead{(kpc$^3$ yr)}   }
\startdata
{\underline{$n_{\rm{o}} = 0.5$ cm$^{-3}$}} & & & & \\
halo 65		&    16  &   25,000  &    0.35  &    1500\\
halo 62		&    35  &   18,000  &    0.77  &    980 \\
halo 64		&    72  &   29,000  &    2.2   &   1900 \\ \hline

{\underline{$n_{\rm{o}} = 0.2$ cm$^{-3}$}} & & & & \\
halo 59		&   81   &   61,000  &    2.7   &   4700\\
halo 57		&   81   &   31,000  &    2.7   &   2100\\
halo 58		&   110  &   58,000  &    4.2   &   4800\\ \hline

{\underline{$n_{\rm{o}} = 0.1$ cm$^{-3}$}} & & & & \\
halo 92		&  360   &   110,000  &    18   &   10,000\\
halo 91		&  220   &    49,000  &    10   &   3800 \\
halo 90		&  300   &    91,000  &    15   &   8700 \\ \hline

{\underline{$n_{\rm{o}} = 0.05$ cm$^{-3}$}} & & & & \\
halo 110	&  960   &   170,000  &    63   &   19,000\\
halo 111	&  500   &    75,000  &    29   &   6600 \\
halo 109	&  1300  &   140,000  &    100  &   15,000\\ \hline

{\underline{$n_{\rm{o}} = 0.02$ cm$^{-3}$}} & & & & \\
halo 100	&  740   &   300,000  &    53   &   40,000\\
halo 101	&  750   &   140,000  &    54   &   15,000\\
halo 99		&  1000  &   260,000  &    83   &   34,000\\ \hline

{\underline{$n_{\rm{o}} = 0.01$ cm$^{-3}$}} & & & & \\
halo 103	& 1000   &   480,000  &    84   &   72,000\\
halo 104	& 1000   &   250,000  &    87   &   31,000\\
halo 102	& 1400   &   460,000  &    130  &   69,000\\ \hline

{\underline{$n_{\rm{o}} = 0.005$ cm$^{-3}$}} & & & & \\
halo 106	&  6800   &   800,000  &    940  &   140,000 \\
halo 107 	&  1400   &   500,000  &    140  &   73,000  \\
halo 105	&  1800   &   900,000  &    210  &   160,000 \\ \hline
\enddata
\label{table:sizepersnr}
\end{deluxetable}

\clearpage

\begin{deluxetable}{ccccccc}
\tablewidth{0pt}
\tabletypesize{\footnotesize}
\tablehead{
\colhead{Simulated}     
& \colhead{$\sum_0^{psf}$ \oxyfive\ $\Delta t$}
& \colhead{$\sum$ \oxyfive\ $\Delta t$}
& \colhead{$\sum_0^{psf}$ \nitfour\ $\Delta t$}
& \colhead{$\sum$ \nitfour\ $\Delta t$}
& \colhead{$\sum_0^{psf}$ \carthree\ $\Delta t$}
& \colhead{$\sum$ \carthree\ $\Delta t$} \\
\colhead{Remnant}     
& \colhead{(\oxyfive\ ion yr)}
& \colhead{(\oxyfive\ ion yr)} 
& \colhead{(\nitfour\ ion yr)}
& \colhead{(\nitfour\ ion yr)} 
& \colhead{(\carthree\ ion yr)}
& \colhead{(\carthree\ ion yr)} }
\startdata
halo 65		& 2.7 $\times 10^{58}$ & 9.2 $\times 10^{60}$ & 1.8 $\times 10^{57}$ & 1.0 $\times 10^{60}$ & 1.4 $\times 10^{57}$ & 2.4 $\times 10^{60}$ \\
halo 62		& 2.1 $\times 10^{59}$ & 8.5 $\times 10^{60}$ & 1.6 $\times 10^{58}$ & 9.0 $\times 10^{59}$ & 6.6 $\times 10^{58}$ & 1.6 $\times 10^{60}$ \\
halo 64		& 5.0 $\times 10^{59}$ & 1.9 $\times 10^{61}$ & 4.8 $\times 10^{58}$ & 1.9 $\times 10^{60}$ & 5.1 $\times 10^{58}$ & 3.2 $\times 10^{60}$ \\ \hline
halo 59		& 2.7 $\times 10^{59}$ & 1.9 $\times 10^{61}$ & 1.1 $\times 10^{58}$ & 2.2 $\times 10^{60}$ & 8.6 $\times 10^{57}$ & 4.6 $\times 10^{60}$ \\
halo 57		& 3.2 $\times 10^{59}$ & 2.0 $\times 10^{61}$ & 1.2 $\times 10^{58}$ & 1.6 $\times 10^{60}$ & 1.5 $\times 10^{58}$ & 2.7 $\times 10^{60}$ \\
halo 58		& 2.0 $\times 10^{59}$ & 4.2 $\times 10^{61}$ & 1.2 $\times 10^{58}$ & 3.3 $\times 10^{60}$ & 1.5 $\times 10^{58}$ & 5.4 $\times 10^{60}$ \\ \hline
halo 92		& 1.6 $\times 10^{60}$ & 3.4 $\times 10^{61}$ & 9.9 $\times 10^{58}$ & 3.8 $\times 10^{60}$ & 1.6 $\times 10^{59}$ & 7.5 $\times 10^{60}$ \\
halo 91		& 6.4 $\times 10^{59}$ & 3.6 $\times 10^{61}$ & 6.6 $\times 10^{58}$ & 2.6 $\times 10^{60}$ & 6.1 $\times 10^{58}$ & 4.7 $\times 10^{60}$ \\
halo 90		& 8.1 $\times 10^{59}$ & 7.5 $\times 10^{61}$ & 3.4 $\times 10^{58}$ & 5.0 $\times 10^{60}$ & 3.7 $\times 10^{58}$ & 9.0 $\times 10^{60}$ \\ \hline
halo 110	& 2.8 $\times 10^{60}$ & 6.4 $\times 10^{61}$ & 2.9 $\times 10^{59}$ & 6.7 $\times 10^{60}$ & 4.7 $\times 10^{59}$ & 1.3 $\times 10^{61}$ \\
halo 111	& 1.4 $\times 10^{60}$ & 6.0 $\times 10^{61}$ & 8.7 $\times 10^{58}$ & 4.3 $\times 10^{60}$ & 2.4 $\times 10^{59}$ & 8.9 $\times 10^{60}$ \\
halo 109	& 4.5 $\times 10^{60}$ & 1.3 $\times 10^{62}$ & 3.0 $\times 10^{59}$ & 8.3 $\times 10^{60}$ & 5.1 $\times 10^{59}$ & 1.6 $\times 10^{61}$ \\ \hline
halo 100	& 1.3 $\times 10^{60}$ & 1.4 $\times 10^{62}$ & 7.7 $\times 10^{58}$ & 1.4 $\times 10^{61}$ & 9.8 $\times 10^{58}$ & 2.8 $\times 10^{61}$ \\
halo 101	& 1.7 $\times 10^{60}$ & 1.2 $\times 10^{62}$ & 1.5 $\times 10^{59}$ & 8.9 $\times 10^{60}$ & 1.8 $\times 10^{59}$ & 2.1 $\times 10^{61}$ \\
halo 99		& 1.8 $\times 10^{60}$ & 2.4 $\times 10^{62}$ & 1.1 $\times 10^{59}$ & 1.7 $\times 10^{61}$ & 1.3 $\times 10^{59}$ & 3.9 $\times 10^{61}$ \\ \hline
halo 103        & 1.3 $\times 10^{60}$ & 2.4 $\times 10^{62}$ & 8.1 $\times 10^{58}$ & 2.3 $\times 10^{61}$ & 1.2 $\times 10^{59}$ & 5.2 $\times 10^{61}$\\
halo 104        & 1.7 $\times 10^{60}$ & 1.9 $\times 10^{62}$ & 1.3 $\times 10^{59}$ & 1.5 $\times 10^{61}$ & 1.7 $\times 10^{59}$ & 3.7 $\times 10^{61}$\\
halo 102        & 1.9 $\times 10^{60}$ & 3.9 $\times 10^{62}$ & 1.3 $\times 10^{59}$ & 2.9 $\times 10^{61}$ & 1.8 $\times 10^{59}$ & 7.2 $\times 10^{61}$\\ \hline
halo 106	& 8.3 $\times 10^{60}$ & 3.9 $\times 10^{62}$ & 6.7 $\times 10^{59}$ & 3.7 $\times 10^{61}$ & 2.8 $\times 10^{60}$ & 9.4 $\times 10^{61}$ \\
halo 107	& 2.1 $\times 10^{60}$ & 3.7 $\times 10^{62}$ & 1.5 $\times 10^{59}$ & 3.3 $\times 10^{61}$ & 2.4 $\times 10^{59}$ & 8.6 $\times 10^{61}$ \\
halo 105	& 2.6 $\times 10^{60}$ & 7.6 $\times 10^{62}$ & 1.9 $\times 10^{59}$ & 6.4 $\times 10^{61}$ & 3.0 $\times 10^{59}$ & 1.7 $\times 10^{62}$ \\ \hline
\enddata
\label{table:highionspersnr} 
\end{deluxetable}

\clearpage

\begin{deluxetable}{ccccccc}
\tablewidth{0pt}
\tabletypesize{\footnotesize}
\tablehead{
\colhead{Simulated}     
& \colhead{$\sum_0^{psf} L_{O^{+5}} \ \Delta t$}
& \colhead{$\sum L_{O^{+5}} \ \Delta t$}
& \colhead{$\sum_0^{psf} L_{N^{+4}} \ \Delta t$}
& \colhead{$\sum L_{N^{+4}} \ \Delta t$}
& \colhead{$\sum_0^{psf} L_{C^{+3}} \ \Delta t$}
& \colhead{$\sum L_{C^{+3}} \ \Delta t$} \\
\colhead{Remnant}     
& \colhead{(ergs)}
& \colhead{(ergs)} 
& \colhead{(ergs)}
& \colhead{(ergs)} 
& \colhead{(ergs)}
& \colhead{(ergs)} }
\startdata
halo 65		&5.6 $\times 10^{47}$ & 1.0 $\times 10^{49}$ & 5.4 $\times 10^{46}$ & 1.8 $\times 10^{48}$	& 4.6 $\times 10^{46}$ &  7.4 $\times 10^{48}$	\\
halo 62		&3.2 $\times 10^{48}$ & 1.0 $\times 10^{49}$ & 2.4 $\times 10^{48}$ & 7.4 $\times 10^{48}$	& 2.4 $\times 10^{48}$ &  7.4 $\times 10^{48}$	\\
halo 64		&1.4 $\times 10^{49}$ & 3.9 $\times 10^{49}$ & 2.1 $\times 10^{48}$ & 7.1 $\times 10^{48}$	& 3.0 $\times 10^{48}$ &  1.4 $\times 10^{49}$	\\ \hline
halo 59		&3.3 $\times 10^{48}$ & 1.4 $\times 10^{49}$ & 1.6 $\times 10^{47}$ & 1.8 $\times 10^{48}$	& 1.3 $\times 10^{47}$ &  5.0 $\times 10^{48}$	\\
halo 57		&3.9 $\times 10^{48}$ & 1.6 $\times 10^{49}$ & 1.7 $\times 10^{47}$ & 1.6 $\times 10^{48}$	& 2.2 $\times 10^{47}$ &  4.4 $\times 10^{48}$	\\
halo 58		&2.1 $\times 10^{48}$ & 2.8 $\times 10^{49}$ & 1.4 $\times 10^{47}$ & 3.4 $\times 10^{48}$	& 1.8 $\times 10^{47}$ &  1.1 $\times 10^{49}$	\\ \hline
halo 92		&9.4 $\times 10^{48}$ & 1.6 $\times 10^{49}$ & 7.6 $\times 10^{47}$ & 1.9 $\times 10^{48}$	& 1.1 $\times 10^{48}$ &  4.6 $\times 10^{48}$	\\
halo 91		&3.4 $\times 10^{48}$ & 1.5 $\times 10^{49}$ & 5.2 $\times 10^{47}$ & 2.2 $\times 10^{48}$	& 5.8 $\times 10^{47}$ &  4.2 $\times 10^{48}$	\\
halo 90		&4.6 $\times 10^{48}$ & 3.0 $\times 10^{49}$ & 2.3 $\times 10^{47}$ & 3.2 $\times 10^{48}$	& 2.5 $\times 10^{47}$ &  9.7 $\times 10^{48}$	\\ \hline
halo 110	&7.6 $\times 10^{48}$ & 1.4 $\times 10^{49}$ & 1.1 $\times 10^{48}$ & 2.2 $\times 10^{48}$	& 1.6 $\times 10^{48}$ &  4.2 $\times 10^{48}$	\\
halo 111	&3.6 $\times 10^{48}$ & 1.5 $\times 10^{49}$ & 2.8 $\times 10^{47}$ & 1.6 $\times 10^{48}$	& 9.0 $\times 10^{47}$ &  4.2 $\times 10^{48}$	\\
halo 109	&1.2 $\times 10^{49}$ & 3.4 $\times 10^{49}$ & 1.0 $\times 10^{48}$ & 3.5 $\times 10^{48}$	& 1.4 $\times 10^{48}$ &  6.4 $\times 10^{48}$	\\ \hline
halo 100	&1.2 $\times 10^{48}$ & 1.2 $\times 10^{49}$ & 9.1 $\times 10^{46}$ & 1.6 $\times 10^{48}$	& 1.3 $\times 10^{47}$ &  2.7 $\times 10^{48}$	\\
halo 101	&1.4 $\times 10^{48}$ & 1.9 $\times 10^{49}$ & 1.7 $\times 10^{47}$ & 2.1 $\times 10^{48}$	& 2.6 $\times 10^{47}$ &  3.3 $\times 10^{48}$	\\
halo 99		&1.4 $\times 10^{48}$ & 3.5 $\times 10^{49}$ & 1.0 $\times 10^{47}$ & 3.5 $\times 10^{48}$	& 1.4 $\times 10^{47}$ &  5.0 $\times 10^{48}$	\\ \hline
halo 103 	&4.0 $\times 10^{47}$ & 1.2 $\times 10^{49}$ & 3.1 $\times 10^{46}$ & 1.6 $\times 10^{48}$	& 5.7 $\times 10^{46}$ &  3.0 $\times 10^{48}$	\\
halo 104 	&5.1 $\times 10^{47}$ & 2.0 $\times 10^{49}$ & 5.0 $\times 10^{46}$ & 1.9 $\times 10^{48}$	& 8.6 $\times 10^{46}$ &  3.0 $\times 10^{48}$	\\
halo 102 	&5.1 $\times 10^{47}$ & 3.8 $\times 10^{49}$ & 4.1 $\times 10^{46}$ & 3.6 $\times 10^{48}$	& 7.1 $\times 10^{46}$ &  5.6 $\times 10^{48}$	\\ \hline
halo 106	&1.1 $\times 10^{48}$ & 1.5 $\times 10^{49}$ & 1.3 $\times 10^{47}$ & 1.9 $\times 10^{48}$	& 6.3 $\times 10^{47}$ &  4.6 $\times 10^{48}$	\\
halo 107	&2.2 $\times 10^{47}$ & 2.2 $\times 10^{49}$ & 2.1 $\times 10^{46}$ & 2.3 $\times 10^{48}$	& 4.1 $\times 10^{46}$ &  4.3 $\times 10^{48}$	\\
halo 105	&2.4 $\times 10^{47}$ & 4.5 $\times 10^{49}$ & 2.1 $\times 10^{46}$ & 4.4 $\times 10^{48}$	& 4.2 $\times 10^{46}$ &  8.8 $\times 10^{48}$ 	\\ \hline
\enddata
\label{table:highionsemisspersnr}
\end{deluxetable}

\clearpage

\begin{deluxetable}{ccc}
\tablewidth{0pt}
\tablecaption{}
\tablehead{
\colhead{Simulated}     
& \colhead{$\sum_0^{psf}{L_{1/4\ keV}} \times \Delta t$}
& \colhead{$\sum {L_{1/4\ keV}} \times \Delta t$} \\
\colhead{Remnant}     
& \colhead{(ROSAT R1 + R2 counts cm$^2$)}
& \colhead{(ROSAT R1 + R2 counts cm$^2$)}   }
\startdata
halo 65		& 6.9 $\times 10^{59}$  & 2.4 $\times 10^{60}$ \\
halo 62		& 2.2 $\times 10^{60}$  & 5.1 $\times 10^{60}$ \\
halo 64		& 9.9 $\times 10^{59}$  & 3.1 $\times 10^{60}$ \\ \hline
halo 59		& 8.4 $\times 10^{59}$  & 1.7 $\times 10^{60}$ \\
halo 57		& 8.0 $\times 10^{59}$  & 1.7 $\times 10^{60}$ \\
halo 58		& 1.4 $\times 10^{60}$  & 4.0 $\times 10^{60}$ \\ \hline
halo 92		& 9.5 $\times 10^{59}$  & 1.2 $\times 10^{60}$ \\
halo 91		& 7.2 $\times 10^{59}$  & 1.3 $\times 10^{60}$ \\
halo 90		& 1.3 $\times 10^{60}$  & 3.0 $\times 10^{60}$ \\ \hline
halo 110	& 7.0 $\times 10^{59}$  & 9.0 $\times 10^{59}$ \\
halo 111	& 5.1 $\times 10^{59}$  & 1.1 $\times 10^{60}$ \\
halo 109	& 1.3 $\times 10^{60}$  & 2.5 $\times 10^{60}$ \\ \hline
halo 100	& 2.6 $\times 10^{59}$  & 7.2 $\times 10^{59}$ \\
halo 101	& 2.3 $\times 10^{59}$  & 1.2 $\times 10^{60}$ \\
halo 99		& 4.1 $\times 10^{59}$  & 2.9 $\times 10^{60}$ \\ \hline
halo 103 	& 1.5 $\times 10^{59}$  & 5.9 $\times 10^{59}$ \\
halo 104 	& 1.3 $\times 10^{59}$ & 1.5 $\times 10^{60}$ \\
halo 102 	& 2.3 $\times 10^{59}$ & 3.5 $\times 10^{60}$ \\ \hline
halo 106	& 2.1 $\times 10^{59}$  & 5.8 $\times 10^{59}$ \\
halo 107	& 7.4 $\times 10^{58}$  & 1.5 $\times 10^{60}$ \\
halo 105	& 1.2 $\times 10^{59}$  & 3.7 $\times 10^{60}$ \\ \hline
\enddata
\label{table:xrayspersnr}
\end{deluxetable}

\clearpage

\subsection{SNR Population: Area Coverage and Volume Occupation}

In this and the following two subsections, we predict the effects of
the population of isolated extra-planar supernova remnants.
For example, we will calculate the fraction of the high latitude
sky's area that is covered by supernova remnants residing
above a height, $\mathcal{Z}$, of 130 pc.
The necessary integral, presented in Section 3.1, will
be approximated by the summation
$\sum_{\mathcal{Z}}^\infty ({\rm{Rate}}_{\rm{SN}}(z) \times
\sum \pi R^2(z,t) \Delta t)$.
The first step in handling the gradients in ambient density and SN rate
with respect to $z$
is to segment the thick disk and halo into six
plane parallel slabs whose boundaries reside at
the midpoints between $z_1$, $z_2$, $z_3$,$z_4$, $z_5$, $z_6$, and $z_7$. 
The top slab extends to $z = \infty$. 
(This parceling scheme begins at $z = 130$~pc, the scaleheight
for the Galactic \hone\ layer, thus excludes
the $z_1$, $n_{\rm{o}} = 0.5$~cm$^{-3}$ simulations which had been created 
for comparison with the other simulations.)

We then integrate Rate$_{\rm{SN}}(z)$, 
the isolated SN rate per unit volume and time given in
Equation~\ref{eq:snrate}, from the lower to the upper boundary 
of each slab.  We multiply each slab's rate integral 
by the time integrated cross sectional area for a single SNR 
residing in that slab
tabulated in Table~\ref{table:sizepersnr}.   This act
implicitly treats all SNRs residing within a slab as if they are residing
at the
representative height of the slab (i.e. $z_2$, $z_3$, etc.).  
For each of the slabs, 
Table~\ref{fraction} lists the integrated SN rates and 
the fraction of area covered by the population of SNRs if 
the remnants are stationary.
Next, we calculate the effects of the
entire population of isolated SNRs born above 130 pc.
We sum the fractional area coverages
of the six slabs.  
For stationary SNRs, we use the area coverage values listed in
Table~\ref{fraction}.
For buoyant SNRs, we adjust the time integrated cross sectional
areas to account for the time spent in the various slabs.  
The results for both buoyant and non-buoyant SNRs are tabulated in
Table~\ref{fractionbuoyant}.
Note that we maintain
accuracy to several significant digits during all calculations, but
round to two significant digits when reporting results.

Table~\ref{fraction} implies that about $1\%$ of the high latitude
sky is covered by young, presumably bright and recognizable SNRs
(the pre-shell formation remnants).
Depending on the assumed simulation parameters and drag coefficient,
$30\%$ to $90\%$ of the sky is covered by the hot gas in SNRs of any 
age\footnote{The cross section of a SNR residing at a given height may 
overlap with 
the cross section of a SNR residing at another height.  Therefore 
the probability that one or more SNR lie along a sightline
perpendicular to the Galactic plane is somewhat smaller than the
sums presented.}.
Although the older SNRs that account for most of the area coverage
are not as recognizable as their
younger cousins, they still harbor large stores of \oxyfive,
\nitfour, and \carthree\ which are observationally important.

The ``filling fraction'', 
the fraction of volume at a given height that is
filled by extraplanar supernova remnants is simply: 
${\rm{Rate}}_{\rm{SN}}(z) \times \sum \frac{4}{3} \pi R^3(z,t) \ \Delta t$.
We calculate the filling fraction at the representative height of
a slab (i.e. $z_2$, $z_3$, $z_4$, $z_5$, $z_6$, or $z_7$) 
by calculating the SN explosion rate at the chosen height
and then multiplying by the time integrated volume occupation
of a single SNR born at the chosen height (taken from 
Table~\ref{table:sizepersnr}).  
The resulting filling fractions for non-buoyant SNRs
are listed in Table~\ref{fraction}.
At all examined heights, isolated non-buoyant SNR bubbles 
fill less than $10\%$ of the volume.
If the remnants are buoyant, then
the filling fractions in the first couple of slabs would
be even smaller, while the minute filling fractions in the upper slabs
would be somewhat larger.
The smallness of our estimates
echo those from other studies, such as \citet{ferriere98-2}.
On account of their small volume occupations, it is very unlikely that
hot supernova remnant bubbles will collide with each other.
Interestingly, the volume filling fraction reaches its maximum
near the SN scaleheight.  This is because
the SN rate decreases as a function of $z$, the 
remnants' size and longevity increase as a function of $z$,
and the maximum product occurs around $z = 300$~pc.

\clearpage

\begin{deluxetable}{ccccccc}
\tablewidth{0pt}
\tabletypesize{\scriptsize}
\tablehead{
\colhead{Slab}     
&\colhead{Height}     
& \colhead{$\int$SN Rate\ $dz$} 
& \colhead{PSF Area}
& \colhead{Total Area}
& \colhead{PSF Volume} 
& \colhead{Total Volume} \\
\colhead{}     
&\colhead{(pc)}     
& \colhead{(kpc$^{-2}$yr$^{-1}$)} 
& \colhead{Coverage ($\%$)} 
& \colhead{Coverage ($\%$)} 
& \colhead{Filled ($\%$)}
& \colhead{Filled ($\%$)}} 
\startdata
$E_o = 0.5 \times 10^{51}$ ergs         & & & & & & \\
{\underline{$B_{eff} = 2.5\ \mu$ G}}    & & & & & & \\
$n$ = 0.2 atom cm$^{-3}$ & 130 to 250   & 1.5 $\times 10^{-6}$ & 
0.012   &   8.9 &  0.0034 &     6.0 \\
$n$ = 0.1 atom cm$^{-3}$ & 250 to 390   & 1.15 $\times 10^{-6}$ & 
0.041   &   12 &   0.014 &     7.9 \\
$n$ = 0.05 atom cm$^{-3}$ & 390 to 670   & 1.0 $\times 10^{-6}$ &  0.099   &   18 &   0.024 &     7.2 \\
$n$ = 0.02 atom cm$^{-3}$ & 670 to 1100  & 4.8 $\times 10^{-7}$ &  0.036   &   15 &  0.0062 &     4.7 \\
$n$ = 0.01 atom cm$^{-3}$ & 1100 to 1600 & 1.2 $\times 10^{-7}$ &  0.012   &   5.6 &  0.0020 &     1.7 \\
{\underline{$n$ = 0.005 atom cm$^{-3}$}} & {\underline{1600 to $\infty$}} & {\underline{2.5 $\times 10^{-8}$}} & {\underline{0.017}}   &   {\underline{2.0}} & 0.00028 &   0.041 \\
Sum & 130 to $\infty$ & 4.3 $\times 10^{-6}$ & 0.22 & 61 &  &  \\ 
 & & & & & \\ \hline

$E_o = 0.5 \times 10^{51}$ ergs         & & & & & & \\
{\underline{$B_{eff} = 5.0\ \mu$ G}}    & & & & & & \\
$n$ = 0.2 atom cm$^{-3}$ & 130 to 250   & 1.5 $\times 10^{-6}$ & 0.012   &   4.6 &  0.0034 &     2.7 \\
$n$ = 0.1 atom cm$^{-3}$ & 250 to 390   & 1.2 $\times 10^{-6}$ & 0.026   &   5.6 &  0.0078 &     3.0 \\
$n$ = 0.05 atom cm$^{-3}$ & 390 to 670   & 1.0 $\times 10^{-6}$ & 0.051  &   7.7 &   0.011 &     2.5 \\
$n$ = 0.02 atom cm$^{-3}$ & 670 to 1100  & 4.8 $\times 10^{-7}$ & 0.036   &   6.8 &  0.0063 &     1.7 \\
$n$ = 0.01 atom cm$^{-3}$ & 1100 to 1600 & 1.2 $\times 10^{-7}$ & 0.012   &   3.0 &  0.0021 &    0.73 \\
{\underline{$n$ = 0.005 atom cm$^{-3}$}} & {\underline{1600 to $\infty$}} & {\underline{2.5 $\times 10^{-8}$}} & {\underline{0.0035}}   &   {\underline{1.3}} & $4.2 \times 10^{-5}$ &   0.022 \\
Sum & 130 to $\infty$ & 4.3 $\times 10^{-6}$ & 0.14 & 29 &  &  \\ 

 & & & & & \\ \hline
$E_o = 1.0 \times 10^{51}$ ergs         & & & & & & \\
{\underline{$B_{eff} = 5.0\ \mu$ G}}    & & & & & & \\ 
$n$ = 0.2 atom cm$^{-3}$ & 130 to 250   & 1.5 $\times 10^{-6}$ & 0.016   &   8.4 &  0.0053 &     6.0 \\
$n$ = 0.1 atom cm$^{-3}$ & 250 to 390   & 1.2 $\times 10^{-6}$ & 0.034   &   11 &   0.012 &     6.8 \\
$n$ = 0.05 atom cm$^{-3}$ & 390 to 670   & 1.0 $\times 10^{-6}$ & 0.14   &   14 &   0.038 &     5.8 \\
$n$ = 0.02 atom cm$^{-3}$ & 670 to 1100  & 4.8 $\times 10^{-7}$ & 0.048   &   12.5 &  0.0097 &     3.9 \\
$n$ = 0.01 atom cm$^{-3}$ & 1100 to 1600 & 1.2 $\times 10^{-7}$ & 0.016   &   5.4 &  0.0031 &     1.6 \\
{\underline{$n$ = 0.005 atom cm$^{-3}$}} & {\underline{1600 to $\infty$}} & {\underline{2.5 $\times 10^{-8}$}} & {\underline{0.0046}}   &   {\underline{2.3}} & $6.3 \times 10^{-5}$ &   0.049 \\
Sum & 21 to $\infty$ & 6.4 $\times 10^{-6}$ & 0.25 & 54 &  &  \\ 
 & & & & & \\ \hline
\enddata
\label{fraction}
\end{deluxetable}

\clearpage

\begin{deluxetable}{cccc}
\tablewidth{0pt}
\tablecaption{ 
Area covered by non-buoyant or buoyant SNR bubbles.
Case 1 assumes $E_o = 0.5 \times 10^{51}$ ergs and $B_{eff} = 2.5\ \mu$G.
Case 2 assumes $E_o = 0.5 \times 10^{51}$ ergs and $B_{eff} = 5.0\ \mu$G.
Case 3 assumes $E_o = 1.0 \times 10^{51}$ ergs and $B_{eff} = 5.0\ \mu$G
}
\tablehead{
\colhead{Case}     
&\colhead{Coverage ($\%$)}     
&\colhead{Coverage ($\%$)}     
& \colhead{Coverage ($\%$)} \\
\colhead{}
&\colhead{(stationary)}
&\colhead{(drag coefficient = 1)}
&\colhead{(drag coefficient = 0)}}
\startdata
1 & 61 & 65 & 87  \\

2 & 29 & 29 & 32  \\ 

3 & 54 & 55 & 67 \\ \hline

\enddata
\label{fractionbuoyant}
\end{deluxetable}

\clearpage


\subsection {The \carthree, \nitfour, and \oxyfive\ Predictions for the
Population of SNRs}

Here, we estimate the time and space averaged column densities of
collisionally ionized \oxyfive, \nitfour, and \carthree\ ions along
high latitude lines of sight due to the population of isolated extraplanar 
SNRs.
The column density of an ion residing above a height of $\mathcal{Z}$ is 
$\int_\mathcal{Z}^\infty ({\rm{Rate}}_{\rm{SN}}(z) \times 
\int \int_0^R n_{\rm{ion}}(r,t) \ 4 \pi r^2 \ dr \ dt) \ dz$.
We replace the integrals with summations.
For each slab, we integrate the supernova explosion rate with respect to
$z$ from the bottom to the top of the slab
(see Table~\ref{fraction}), and multiply by the 
\oxyfive, \nitfour, and \carthree\ contributions made by a SNR born within
the slab (see Table~\ref{table:highionspersnr} for stationary SNRs).  
When calculating the \oxyfive, \nitfour, and \carthree\ 
contributions made by buoyant SNRs,
we take into account the extended lifetime of the SNR and the
time spent in more rarefied environments.
We then sum across all slabs above $z = 130$ pc.
Table~\ref{highioncoldenpopulation} tabulates the resulting
average column densities, for remnants that are assumed
to be buoyant as well as those that are not.
For example, the high latitude sky average \oxyfive\ column density ranges
from 2.3 to $6.3 \times 10^{13}$ cm$^{-2}$, depending on the
simulation parameters and drag coefficient.
This range may appear to be quite compact.
Moderately buoyant SNRs would produce only 3 to $20\%$ more
\oxyfive\ than stationary SNRs, while
unfettered SNRs would produce 35 to $110\%$ more \oxyfive\
than stationary SNRs.
For identical drag coefficients, Cases 1, 2, and 3
vary by a factor of 2.1 to 2.4, primarily due to the factor
of 2 variation in assumed explosion energy.   Interestingly,
increasing the nonthermal pressure by a factor of 4 (from Case 1
to Case 2) slightly {\underline{decreases}} the net \oxyfive\ contribution.
The reason for this phenomenon is that
remnants experiencing greater nonthermal pressure are smaller,
therefore denser, therefore more luminous, and therefore 
shorter lived.
Note that Table~\ref{highioncoldenpopulation}'s values were found 
by averaging over many SNRs
and even parts of the sky lacking SNRs.   For predictions of
\oxyfive, \nitfour, and \carthree\ column densities {\it{within}}
individual SNRs, as a function of age and radius,
see Paper I.   

\begin{figure}           
  \plotone{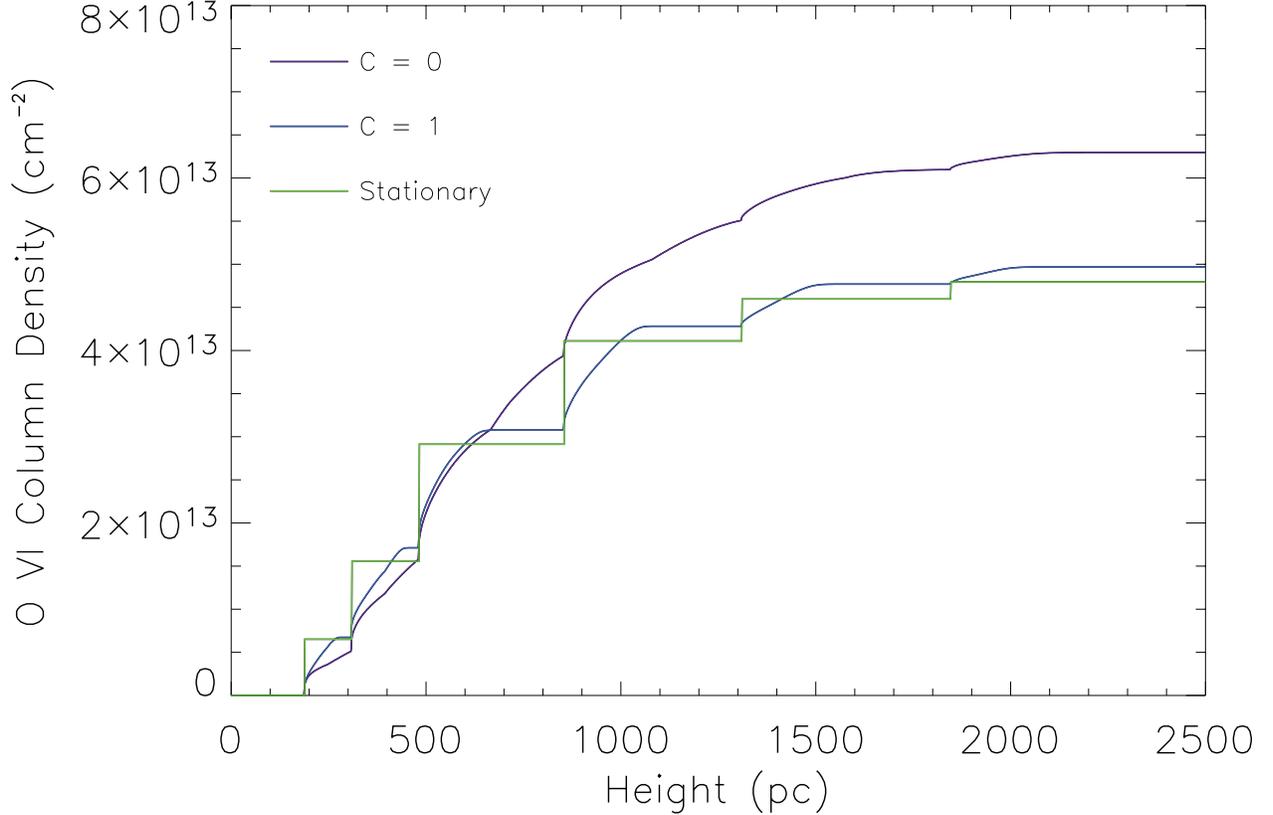}
  \caption[Column Densities due to stationary and buoyant SNRs]
{The \oxyfive\ column density expected to reside on
vertical sightlines extending from the Earth to given
heights above the plane, as a function of height above the plane.
Various choices of buoyancy are marked by color.
Green denotes a population of stationary SNRs.
Blue denotes a population of moderately buoyant SNRs (drag coefficient = 1).
Purple denotes a population of unrestrainedly buoyant SNRs 
(drag coefficient = 0).
Independent of the choice of drag coefficient, 
the $z = 390$ to 670~pc slab makes the greatest contribution
to the total column density through the halo.
Note that the \oxyfive\ {\it{appears}} to begin at $z = 190$~pc, only
because we set off the lowest SN explosions at the representative
height for the lowest included slab, which is $z_2 = 190$~pc.
As in Figure~\ref{fig:buoyancy},
the simulated SNRs have an explosion energy of $1.0 \times 10^{51}$~ergs
and ambient effective magnetic field of $5.0 \mu$G (i.e. Case 3).
}
\label{fig:colden_vs_z}
\end{figure}

The height distribution of the \oxyfive-rich gas in extraplanar
SNRs is illustrated in Figure~\ref{fig:colden_vs_z}.
For all choices of simulation parameters and drag coefficients,
the $z = 390$ to 670~pc slab
makes the largest contribution to the sky averaged \oxyfive\ column
density.   
In Section 4.2, we will add the local \oxyfive\ distribution
and compare the result with \copernicus, \orfeus, and \fuse\ observations.

\begin{figure}           
  \plottwo{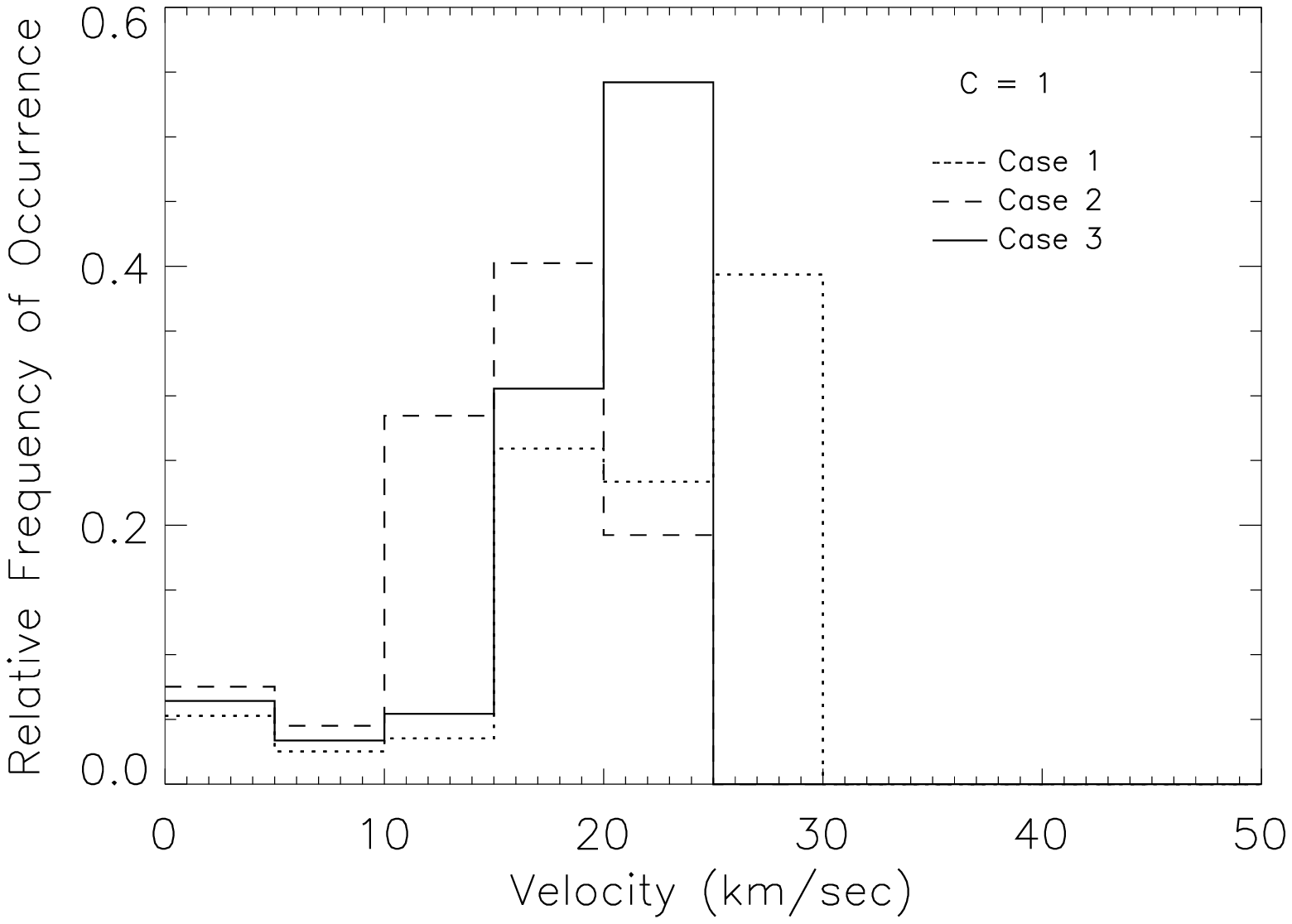}{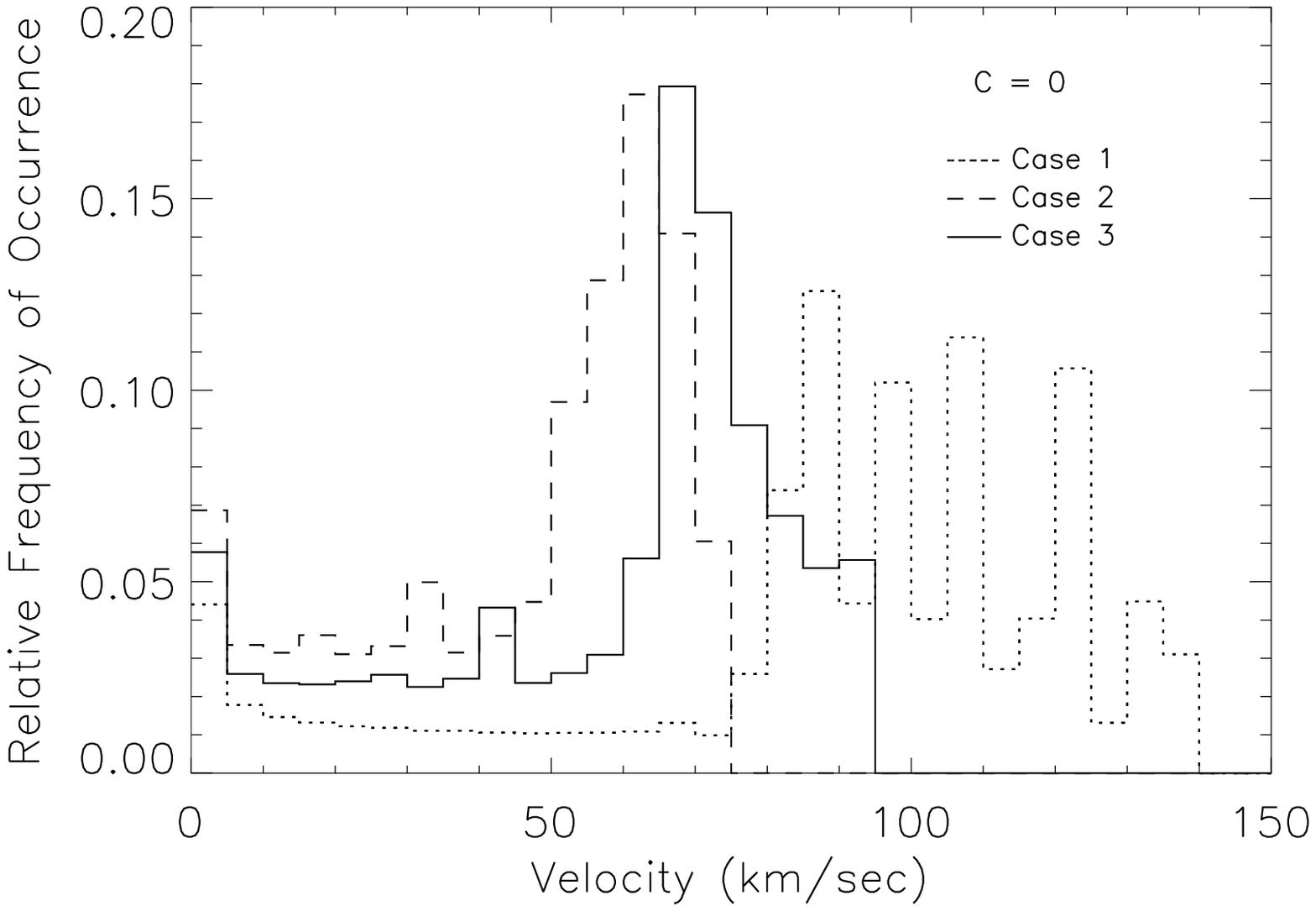}
  \caption[Velocity Distribution for Buoyant SNRs]
{Distribution in buoyant velocity among the population of extraplanar
supernova remnants, 
assuming moderate drag (C = 1, left figure) or no drag 
(C = 0, right figure). In addition to buoyant bulk velocity directed
away from the Galactic plane,
the young remnants will also exhibit expansion velocities.  
Detailed predictions of the expansion are provided in Paper I.}
\label{fig:velocity}
\end{figure}

Figure~\ref{fig:velocity} displays the distribution of buoyant
velocities for the population of supernova remnants.  In making
these plots, we assumed that the drag coeffient is 0 or 1.   
No plot is needed for the stationary SNRs, because their buoyant 
velocities are 0 km s$^{-1}$.
The plotted buoyant velocities are directed
away from the Galactic plane.   The first step in calculating the
distributions was to calculate each remnant's buoyant velocity
as a function of time by integrating the buoyant acceleration 
with respect to time using small time steps.  
We assumed that buoyancy does not tear the remnants apart,
so the calculated buoyant velocity was assumed to pertain to the 
entire supernova remnant.
Remnants born at different heights above the plane have different
buoyancy histories.   Thus, the final distributions of velocities
were found by making weighted compilations of the remnants'
buoyancy histories, where the weightings were proportional to the
supernova explosion rate in the given slab.
Case 1 ($E_o = 1.0 \times 10^6$~ergs, $B_{eff} = 5 \mu$G)
remnants are the most buoyant, probably because the small
assumed ambient nonthermal pressure allows the hot bubbles
to expand to greater volumes, lowering the internal density and
raising their buoyancy.  Case 2 remnants are the least buoyant
for the converse reason.  With drag coefficients of 0 and 1, 
we only see the \oxyfive-rich gas
moving away from the plane;  we do not see it stall and begin to
fall back toward the plane.
However, we saw the \oxyfive-rich gas rise and fall in
preliminary estimates using larger drag coefficients.
If there is moderate drag ($C = 1$)
then the typical buoyant velocity will hover around 20 km sec$^{-1}$.
If there is no drag (C = 0), then the typical buoyant velocity increases 
to $\sim70$ to $\sim100$ km sec$^{-1}$.
If we were to observe such remnants, relative motions within the Galaxy 
would probably increase the
spread of observed velocities, as would motion within the remnants themselves
(see Paper I for estimates).
Observations pointed at
non-zenith directions record only the radial velocity,
a trigonometric fraction of the velocity in the vertical direction.

\clearpage

\clearpage

\clearpage

The proceedure for calculating the high ion 
resonance line intensities is very similiar to that for calculating
the column densities. 
Theoretically, the average intensities are given by 
$\int_\mathcal{Z}^\infty ({\rm{Rate}}_{\rm{SN}}(z) \times 
\int L(z,t) dt) \ dz /(4 \pi {\rm{steradians}})$.  
We replace the integrals with summations.   
The resulting intensities are listed in 
Table~\ref{highionintensitypopulation}.   
Allowing unrestrained buoyancy increases
the sky-averaged intensities by only $20\%$ or less, thus
much less significantly than it affects the 
high-stage ion column densities.
Buoyancy mostly affects the very late-term evolution of the remnants
by extending their lives and elevating them into more rarefied 
surroundings.  However, old remnants are dim and moving them to
more rarefied surroundings makes them even dimmer.
Therefore, these two influences on net photon production counteract
each other.
For the same reason, we see that buoyancy has almost no effect
in the following calculations of
$1/4$~keV X-ray intensity.


\clearpage

\begin{deluxetable}{cccc}
\tablewidth{0pt}
\tablecaption{The model population of SNRs in the thick disk produce the
following sky-averaged \oxyfive, \nitfour, and \carthree\ column densities.
Case 1 assumes $E_o = 0.5 \times 10^{51}$ ergs and $B_{eff} = 2.5\ \mu$G.
Case 2 assumes $E_o = 0.5 \times 10^{51}$ ergs and $B_{eff} = 5.0\ \mu$G.
Case 3 assumes $E_o = 1.0 \times 10^{51}$ ergs and $B_{eff} = 5.0\ \mu$G
}
\tablehead{
\colhead{Ion and Case}     
& \colhead{Column Density}
& \colhead{Column Density}
& \colhead{Column Density} \\
\colhead{}
& \colhead{(ions cm$^{-2}$)}
& \colhead{(ions cm$^{-2}$)}
& \colhead{(ions cm$^{-2}$)} \\
\colhead{}     
& \colhead{(stationary SNR)}
& \colhead{(drag coefficient = 1)}
& \colhead{(drag coefficient = 0)} 
}
\startdata
\underline{\oxyfive} &  &  &  \\
1	&	$2.5 \times 10^{13}$	&	$2.7 \times 10^{13}$	&	$4.0 \times 10^{13}$	\\
2	&	$2.3 \times 10^{13}$	&	$2.4 \times 10^{13}$	&	$2.6 \times 10^{13}$	\\
3	&	$4.8 \times 10^{13}$	&	$5.0 \times 10^{13}$	&	$6.3 \times 10^{13}$	\\ \hline
\underline{\nitfour} &  &  &  \\
1	&	$2.6 \times 10^{12}$	&	$2.9 \times 10^{12}$	&	$4.2 \times 10^{12}$ \\
2	&	$1.7 \times 10^{12}$	&	$1.8 \times 10^{12}$	&	$2.1 \times 10^{12}$ \\
3	&	$3.4 \times 10^{12}$	&	$3.5 \times 10^{12}$	&	$4.7 \times 10^{12}$ \\ \hline
\underline{\carthree} &  &  &  \\
1	&	$5.3 \times 10^{12}$	&	$6.3 \times 10^{12}$	&	$1.1 \times 10^{13}$ \\
2	&	$3.7 \times 10^{12}$	&	$3.8 \times 10^{12}$	&	$5.0 \times 10^{12}$ \\
3	&	$7.0 \times 10^{12}$	&	$7.4 \times 10^{12}$	&	$1.2 \times 10^{13}$ \\
\enddata
\label{highioncoldenpopulation}
\end{deluxetable}

\clearpage

\begin{deluxetable}{cccc}
\tablewidth{0pt}
\tablecaption{The model population of SNRs in the thick disk produce the
following sky-averaged \oxyfive, \nitfour, and \carthree\ resonance line
doublet intensities.
Case 1 assumes $E_o = 0.5 \times 10^{51}$ ergs and $B_{eff} = 2.5\ \mu$G.
Case 2 assumes $E_o = 0.5 \times 10^{51}$ ergs and $B_{eff} = 5.0\ \mu$G.
Case 3 assumes $E_o = 1.0 \times 10^{51}$ ergs and $B_{eff} = 5.0\ \mu$G
}
\tablehead{
\colhead{Ion and Case}     
& \colhead{Intensity}
& \colhead{Intensity}
& \colhead{Intensity} \\
\colhead{}
& \colhead{(photon s$^{-1}$ cm$^{-2}$ sr$^{-1}$)}
& \colhead{(photon s$^{-1}$ cm$^{-2}$ sr$^{-1}$)}
& \colhead{(photon s$^{-1}$ cm$^{-2}$ sr$^{-1}$)} \\
\colhead{}     
& \colhead{(stationary SNR)}
& \colhead{(drag coefficient = 1)}
& \colhead{(drag coefficient = 0)} 
}
\startdata
\underline{\oxyfive} &  &  &  \\
1	&	840	&	850	&	920	\\
2	&	940	&	960	&	1000	\\
3	&	1800	&	1900	&	2100 	\\ \hline
\underline{\nitfour} &  &  &  \\
1	&	130	&	140	&	150	\\
2	&	130	&	130	&	140	\\
3	&	240	&	250	&	270	\\ \hline
\underline{\carthree} &  &  &  \\
1	&	390	&	390	&	410	\\
2	&	370	&	370	&	380	\\
3	&	780	&	790	&	830	\\ 
\enddata
\label{highionintensitypopulation}
\end{deluxetable}

\clearpage

\subsection{The X-ray Predictions for the Population of SNRs}

Following the procedures outlined in the previous two subsections,
we estimate the average $\frac{1}{4}$ keV surface brightness
due to the ensemble of extraplanar supernova remnants
%
(see Table~\ref{averagexray}).  
The average $1/4$~keV countrate is 
120 to 320 $\times 10^{-6}$ counts s$^{-1}$ arcmin$^{-2}$,
depending on the choice of explosion energy, magnetic pressure, and
to a minute extent, drag coefficient.   
These estimates would increase by about $50\%$ if we were to add the
$z_1$ SNRs between $z = 21$ and 130~pc.
The predictions are insensitive to buoyancy for the same reason
as the high ion intensities are little affected by buoyancy -
the longer lifetimes of buoyant remnants are counteracted by their
lesser luminosities in old age.

Because we have averaged the SNR populations' brightness over the entire
high latitude sky, the resulting average countrate is less than the 
countrate from individual remnants, especially young remnants.
Roughly half of the soft X-rays are emitted 
during the bright, but brief stage before the remnant forms a cool shell.
The remaining half are emitted later, when the old, dim 
remnants are difficult to identify.   
Observationally, the collective 
emission from unidentifiable remnants covering half of the sky
would be construed as an X-ray background, punctuated by a few bright
arcs tracing the limbs of young remnants.
We expand on this description in the following subsection.

\clearpage

\begin{deluxetable}{cccc}
\tablewidth{0pt}
\tablecaption{
Sky-Averaged $1/4$~keV Countrates (in the ROSAT R1 + R2 bands)
from the Population of SNRs above $z = 130$~pc. 
Case 1 assumes $E_o = 0.5 \times 10^{51}$ ergs and $B_{eff} = 2.5\ \mu$G.
Case 2 assumes $E_o = 0.5 \times 10^{51}$ ergs and $B_{eff} = 5.0\ \mu$G.
Case 3 assumes $E_o = 1.0 \times 10^{51}$ ergs and $B_{eff} = 5.0\ \mu$G
}
\tablehead{
\colhead{Case}     
& \colhead{Countrate}
& \colhead{Countrate}
& \colhead{Countrate} \\
\colhead{}     
& \colhead{($10^{-6}\ \frac{\rm{counts}}{\rm{s}\ \rm{arcmin}^2}$)}
& \colhead{($10^{-6}\ \frac{\rm{counts}}{\rm{s}\ \rm{arcmin}^2}$)}
& \colhead{($10^{-6}\ \frac{\rm{counts}}{\rm{s}\ \rm{arcmin}^2}$)} \\
\colhead{}     
& \colhead{(stationary SNR)}
& \colhead{(drag coefficient = 1)}
& \colhead{(drag coefficient = 0)} \\ 
}
\startdata
1	&	120	&	120	&	120	\\
2	&	130	&	130	&	130	\\
3	&	310	&	310	&	320	\\
\enddata
\label{averagexray}
\end{deluxetable}

\clearpage

\subsection{Spatial Appearance}

The high latitude sky (outside of superbubbles)
should be comprised of three types of regions,
those having bright young remnants seen in projection 
($\sim 0.2\%$ of sky), 
those having dimmer older remnants seen in projection
($\sim 30$ to $\sim 90\%$ of sky), and 
those having no remnants ($\sim 10$ to $\sim70\%$ of sky).  
%
The young remnants should appear edge brightened in 
$\frac{1}{4}$~keV X-ray emission,
\oxyfive, \nitfour, and \carthree\ emission, and in
numbers of \oxyfive, \nitfour, and \carthree\ ions.
For example, 
the limb of a pre-shell formation
remnant evolving in an $n = 0.01$~cm$^{-3}$ environment
emits more than 
3000 $\times 10^{-6}$ $\frac{1}{4}$~keV counts s$^{-1}$ arcmin$^{-2}$.
It would be easily observed.
Remnants evolving nearer to the plane (i.e. at $z < 1300$ pc)
would be even brighter.

As the edge of its hot bubble cools below $\sim 10^{5.7}$~K
a remnant loses its sharply edge-brightened look.  If becomes
slightly edge brightened and then centrally brightened 
(see Papers I and II
for radial profiles of sample SNRs).
Its total luminosity plummets.  
Our sample SNR's 
$\frac{1}{4}$~keV surface brightness 
drops to $<500 \times 10^{-6}$ counts s$^{-1}$ arcmin$^{-2}$.
An old remnant should be difficult to recognize, considering that 
it may be outshone by the unobscured Local Bubble 
($\sim300$ to $\sim800$~$\times 10^{-6}$~counts~s$^{-1}$~arcmin$^{-2}$) 
combined with the extragalactic background 
($\sim400$~$\times 10^{-6}$~counts~s$^{-1}$~arcmin$^{-2}$, de-absorbed).  
If an old
remnant were to be detected in the X-ray regime, 
it should exhibit a soft spectrum and strong recombination edges
(see Paper II for spectra of sample SNRs). 
Rather than being stars of the show, these remnants 
would create mottled soft X-ray and high stage ion backgrounds. 
Note that if \oxyfive\ column density and $1/4$~keV maps 
were made from the SNR simulations, they would probably 
look somewhat dissimilar because of the significantly different
time evolution in these observables.

\section{Comparison between Simulations and Observations}

\subsection{Soft X-rays}

Early, modest angular resolution surveys 
(the Wisconsin Rocket Program, 
McCammon {\it{et al.}} 1983,
\citet{mbsk},
SAS 3,
\citet{mc}
and the HEAO 1 A-2 Low Energy Detectors,
\citet{gnabfk})
detected bright
soft X-ray ($\sim \frac{1}{4}$~keV) emission from every direction on the sky.
The maps show considerable structure and identifiable features 
such as Loop I, the Eridanus
Soft X-ray Enhancement, and the Monogem Ring.
The poles are brighter than the equator and the northern polar region 
is brighter than the southern polar region. 
Most of the observed soft X-rays originate in the
Local Bubble (a $\sim60$~pc radius
region surrounding the sun) and the Milky Way's halo.
Extragalactic sources produce a 
fraction of the average observed total 
\citep{mmsw,sp}.
Small contributions to the flux observed at low latitudes are made by 
pre-radiative phase remnants and Galactic
point sources.
\citep{rabgghmtv}.

The halo and extragalactic soft X-rays are attenuated by the 
layer of diffuse neutral gas in the disk,
so that only $\sim 2/3$rds reach the solar system. 
In directions with especially opaque clouds,
distinct ``shadows'' appear in the X-ray maps.  It was by
observing shadowed regions with 
the R{\"{o}}ntgensatallit Position Sensitive Proportional Counter 
(or ROSAT PSPC), that researchers
first conclusively demonstrated 
the emissivity
of the distant (halo plus extragalactic) component
\citep{smhhs,bm,khm,smbm,wy}.
More recently, 
\citet{seffp,sfks,ks}
extended this type of analysis to the entire sky.
Their estimates for the surface
brightness of the de-absorbed distant component range from 
$\sim400$ to $3000 \times 10^{-6}$ ROSAT 1/4~keV
counts s$^{-1}$ arcmin$^{-2}$, depending on the direction.  
The high end of the quoted range is misleading, because it refers to 
the flux from anomalous objects, such as the Draco region,
and intermediate distance objects, 
such as the North Polar Spur and the Eridion Superbubble, 
which the analysis technique partitioned between the near and far 
components.
Due to the North Polar Spur and Draco, 
the map of the northern sky is brighter and more complex 
than that of the southern sky.  The southern sky provides only 
$\sim400$ to $1000 \times 10^{-6}$ counts~s$^{-1}$~arcmin$^{-2}$, 
(unabsorbed) and has
an average count rate for $b < -65^{\rm{o}}$
of $810 \times 10^{-6}$ counts~s$^{-1}$~arcmin$^{-2}$.  
Because the southern
halo is far less ``polluted'' by anomalous regions, this paper takes
it as the standard.  Next, 
the extragalactic component must be subtracted from the distant component 
to yield the halo flux.  From shadowing
studies of nearby spiral galaxies, the 
1/4~keV extragalactic intensity has been estimated to be
about 30 keV cm$^{-2}$ s$^{-1}$~sr$^{-1}$~keV$^{-1}$, equivalent to
$400 \times 10^{-6}$
counts~s$^{-1}$~arcmin$^{-2}$ in the \rosat\ $\frac{1}{4}$~keV band
(see 
\citet{sp,brw,csmsw}).
The extragalactic surface brightness should be fairly smooth and isotropic.
Subtracting its flux leaves only
$\sim400 \times 10^{-6}$ counts~s$^{-1}$~arcmin$^{-2}$ 
(unabsorbed), 
which is attributable to the Milky Way's southern halo.

In comparison with our simulation results, 
310 $\times 10^{-6}$ counts~s$^{-1}$~arcmin$^{-2}$, or $3/4$ of the
observed emission could come from the population of extraplanar
SNRs if their explosion energies are $1 \times 10^{51}$ ergs,
their surrounding effective magnetic field is 5 $\mu$G, and they are 
stationary.  
Our estimate rises to 400 $\times 10^{-6}$ counts~s$^{-1}$~arcmin$^{-2}$, 
or $100\%$ of the observationally determined surface brightness
if we use \citet{mw}'s isolated massive progenitor
SN rate rather than averaging it with
\citet{ferriere98}'s rate.
Assuming the more modest rate and the full range of explored parameters 
yields estimated brightnesses
of 120 to 320 $\times 10^{-6}$ counts~s$^{-1}$~arcmin$^{-2}$, or 
30$\%$ to 80$\%$  of the observationally determined flux.

Other important characteristics of the X-ray emitting gas include its
vertical extent, degree of spatial non-uniformity, and temperature.
Determining the height of the X-ray emitting plasma is difficult,
but shadowing by low, intermediate, and high velocity clouds provides some
indication.  After comparing the
$1/4$~keV X-ray emission with the \hi\ column densities in and
around the M complex, 
\citet{hmshbmke}
concluding that 
nearly all of the distant emission originates beyond the low velocity clouds
($ z_{cloud} \stackrel{<}{\sim} 200$~pc), while some originates
below and some originates beyond the high velocity clouds
($ 1.5$~kpc~$\stackrel{<}{\sim} z_{cloud} \stackrel{<}{\sim} 4.4$~kpc).
Hence the distribution of extraplanar SNRs is in rough agreement 
with the distribution of X-ray emitting gas.

The eye can easily identify patchiness in the $1/4$~keV survey maps of 
\citet{seffp}.
Quantitatively,
the halo emission
varies by a factor of 2 over angular scales of $\sim20^{\rm{o}}$
\citep{sefmpsstv}.
On a smaller scale, 
\citet{bm} found that 
the emission behind the Draco cloud varies by 1/3 from the northern
to the southern half of the region ($\sim1^{\rm{o}}$).
In the Milky Way's south, the flux is arranged in a fringe of bright
regions northward of $b \sim -45^{\rm{o}}$
(which may be artifacts of the deabsorption analysis), 
a mottled background, and
a couple of noticeably bright regions below $l \sim -45^{\rm{o}}$
(i.e. the regions around $l = 247^{\rm{o}}$, $b = -64^{\rm{o}}$
and $l = 215^{\rm{o}}$, $b = -68^{\rm{o}}$).
We think that these two bright regions must be emitting features rather than 
artifacts of the analysis technique because 
there is no significant spatial correlation between the
distant emission map and the absorption map used in \citet{sefmpsstv}'s 
analysis.  
The shapes of these bright features can be better seen in 
the less smoothed total observed surface brightness map of 
\citet{sefmpsstv}.
This map shows that the larger feature, the
one around $l = 247^{\rm{o}}$, $b = -64^{\rm{o}}$ has a fairly
crisp outer edge and a crescent, edge brightened shape.  

Approximately 1 bright, pre-shell formation phase SNRs is expected 
to reside in the halo of each hemisphere above
$b {\sim} |50^{\rm{o}}|$ and $z \sim 130$~pc.
The feature around
$l = 247^{\rm{o}}$, $b = -64^{\rm{o}}$ 
is a reasonable candidate for a pre-shell formation phase halo SNR.
The \citet{dl} \hone\ survey shows a cloud abutting
the thickest part of the larger crescent.  We were unable to find
additional confirmation in radio synchrotron surveys 
\citep{hssw,aamo}
or the \rosat\ $3/4$~keV soft X-ray survey \citep{sefmpsstv}.
The lack of confirmation 
is not a confirmation of the lack of a remnant for the following reasons.
\citet{cl}
empirically determined that, on average,
the radio brightness of a SNR decreases with $z$.
Correspondingly, our simulations of SNRs in low density environments found that
their shells were not very dense
and so probably not very emissive in synchrotron emission.
In addition, the $3/4$~keV emission from the simulated SNRs wanes far 
earlier than does the $1/4$~keV emission, thus a remnant can be
observed in $1/4$~keV X-rays after it has become too dim
to be observed in $3/4$~keV X-rays.
The smaller feature, which is around
$l = 215^{\rm{o}}$, $b = -68^{\rm{o}}$, also has an edge brightened
crescent shape, though it is
less crisp and less bright than the larger feature.  
Suggesting that this is a SNR would be more speculative.

The next element of study is the mottled background.  
Galactic X-ray emitting
gas covers roughly half of sky above $|b| = 45^{\rm{o}}$ sky in
\citet{seffp}'s $1/4$~keV survey map 
(assuming that the extragalactic sources contribute
400 $\times 10^{-6}$ counts~s$^{-1}$~arcmin$^{-2}$).
This compares well with the fraction of the sky
covered by post shell formation phase SNRs
(30 to 90$\%$).

Finally, we turn to the spectral temperature of the emission.
Several studies of \rosat's low resolution spectra, 
including \citet{ks}'s and
\citet{sfks}'s found that the high latitude extraplanar
emission has two temperature components, a 
$\sim 10^6$~K, non-homogeneously distributed component
and a weaker, $\sim3 \times 10^6$~K, homogeneously distributed component.
Because they are distributed differently, the 
$10^6$~K and the $3 \times 10^6$~K components are thought to
originate in different regions.   
If extraplanar SNRs are producing much of the 
$10^6$~K component, then we would expect their
soft X-ray emission to look like a $10^6$~K spectrum.
This was found to be the case when the simulated spectra 
were treated as real spectra (i.e.
convolved with the \rosat\ response function and fit
to collisional equilibrium spectral models), 
in Paper II.
Note that the SNR plasma was often out of collisional
ionizational equilibrium and as a result, 
the gas's temperature was generally higher or lower than
the temperature calculated from its spectrum.

\subsection{\oxyfive\ Column Densities and Intensities}

Soft X-rays are emitted by plasma whose carbon, nitrogen, and 
oxygen atoms have been ionized to the hydrogen-like and
helium-like levels.
The neighboring ions,
lithium-like ions of carbon, nitrogen, and oxygen
can be observed by absorption or emission
in their strong resonance line doublets in the 
ultraviolet region of the electromagnetic spectrum 
(\oxyfive: 1032, 1038 \AA; \nitfour: 1239, 1243 \AA;
\carthree: 1548, 1551 \AA).
Owing to its large ionization potential,
the observed \oxyfive\ is thought to be produced via
collisional ionization.   
\nitfour, and \carthree\ can be produced
via photoionization or collisional ionization.
In fact, \citet{smh} have already
shown that substantial numbers of extraplanar \cartwo\ ions have
been ionized to \carthree\ by SNR photons.
Our code simulates collisional
ionization and recombination but does not simulation
photoionization.   Therefore we will confine our comparisons
to the soft X-ray and \oxyfive\ data.

\begin{figure}           
  \vspace{-3cm}
  \plotone{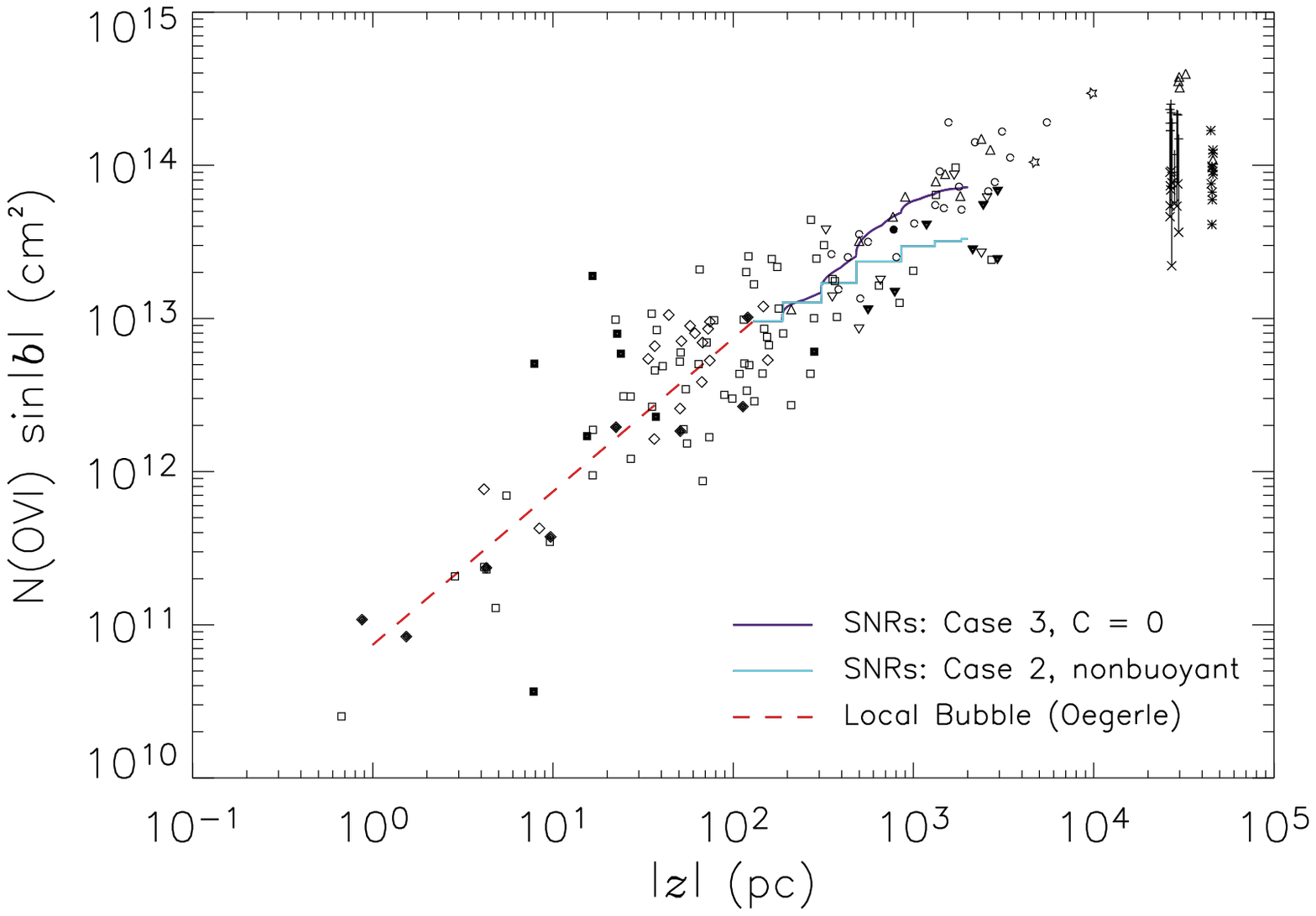}
  \vspace{-1.5cm}
  \caption[SNR simulations versus Copernicus, ORFUES, and FUSE data]
{Observed \oxyfive\ column densities compared with estimates
from the extraplanar SNR simulations (blue and purple curves) combined
with the estimated Local Bubble contribution
(\citet{oegerleetal}, dashed red line).
The purple and blue curves correspond to the least and
most prolific extraplanar SNR models, respectively.
The squares mark \citet{jenkins1978a}'s \copernicus\ observations.
The \orfeus\ observations by \citet{hurwitzbowyer} are marked
with downward pointing triangles while the
\orfeusii\ observations by \citet{widmannetal} are marked
with upward pointing triangles.
The \fuse\ \ Local Bubble observations by \citet{oegerleetal}
are marked with diamonds.
The \fuse\ \ Milky Way halo observations by \citet{zsargoetal}
are marked with circles.
The two \fuse\ \ Milky Way star observations by \citet{wakkeretal03}
are marked with stars.
For the above plotting, detections are marked with open symbols and 
upper limits are marked by solid symbols.
The \citet{howketal} SMC observations are marked with 
asterisks.
The \citet{howketal} LMC observations of \oxyfive\ within
the velocity range $v \leq \pm50$ km s$^{-1}$ are plotted with Xs.
These column densities plus higher positive velocity \oxyfive\
observed along the same sightlines are plotted with pluses.   
All of the \citet{howketal} observations were detections.
Vertical lines connect the higher and lower LMC datapoints.
The agreement between the model and the data
suggests that the extraplanar supernova remnants reasonably
explain the \oxyfive\ observed between $z \sim 130$~pc and $z \sim 2000$~pc.
}  
\label{fig:scaleheight_with_data}
\end{figure}

Hundreds of Milky Way sightlines have been surveyed for \oxyfive\ absorption
(\citet{jenkinsmeloy}; 
\citet{jenkins1978a};
\citet{hurwitzbowyer};
\citet{widmannetal};
\citet{zsargoetal};
\citet{wakkeretal03};
\citet{oegerleetal}).
The data suggest that this ion is
distributed inhomogeneously and is more plentiful nearer to the
plane than at greater distances.
The column density vs height data are plotted in the standard fashion
in Figure~\ref{fig:scaleheight_with_data}.
For comparison, our most productive model
($E_o = 1.0 \times 10^{51}$ ergs, $B_{eff} = 5 \mu$G, drag coefficient = 0)
and least productive model
($E_o = 0.5 \times 10^{51}$ ergs, $B_{eff} = 5 \mu$G, nonbuoyant)
have been plotted.   Because our models begin at $z = 130$~pc,
we have also plotted the expected local contribution (taken from
\citet{oegerleetal}) and added it to our models in order to
obtain the total quantity of \oxyfive\ ions expected to reside
between the Sun and various heights above the plane.
The extraplanar SNR column densities reasonably track the
observations within the relevant height range 
($\sim130 \leq z \leq \sim2000$~pc).
Thus, extraplanar SNRs can explain both the observed column
densities and the increase in column density with height, within
this $z$ range.

The extraplanar SNR model was not meant to explain the
\oxyfive\ observed within the first hundred or so parsecs
of the Sun.   That material
has been attributed to the Local Bubble and clouds within it.   
Nor does the model explain the 
\oxyfive\ residing beyond $ z \sim 2000$~pc (about 6 times the
SN scaleheight).
That column density requires another source.

\clearpage

\clearpage

The upper limits and weak detections in
Figure~\ref{fig:scaleheight_with_data} demonstrate that
the \oxyfive-rich gas does not completely cover the 
high latitude, extraplanar sky.   
Of the sightlines terminating above 130~pc,
$1/8$ have \oxyfive\ upper limits and an additional
$1/8$ have such small column densities
(less \oxyfive\ than expected from local material)
that we can assume they have not encountered extraplanar \oxyfive.
In comparison, our simulation results also imply incomplete
(30 to 90$\%$) sky coverage.

\citet{wakkeretal03}'s Figure 9 displays the 
column density vs velocity distribution of the \oxyfive\ features
observed in their survey.
The distribution consists of a cluster of features with 
$-50$~km~sec$^{-1}$ $\leq v \leq 50$~km~sec$^{-1}$ 
plus separate clusters having 
$-400$~km~sec$^{-1}$ $\leq v \leq -100$~km~sec$^{-1}$ and
$75$~km~sec$^{-1}$ $\leq v \leq 400$~km~sec$^{-1}$.
The high velocity clusters are not symmetric;  the positive high
velocity cluster extends to lower speeds and column densities than does
the negative high velocity cluster.
In comparing with our simulation results, we must keep in mind
that our predicted velocities are in the vertical direction
while \citet{wakkeretal03}'s velocities are along high latitude
lines of sight.    Thus, we would want to multiply the observed 
velocities by cosecant(latititude), where the typical latitude is
around 50$^{\rm{o}}$ to 60$^{\rm{o}}$.
In our calculations, 
the greatest velocity to which the extraplanar SNRs are accelerated is
$140$~km sec$^{-1}$ and this is for only one set of parameter choices.
Such remnants would contribute to
the moderate speed, low column density portion of the observed
$75$~km~sec$^{-1}$ $\leq v \leq 400$~km~sec$^{-1}$ cluster, though they
would not produce the entire cluster or its negative velocity counterpart.
If extraplanar SNRs experience moderate to large drag, then 
their \oxyfive\ features would contribute to the
$-50$~km~sec$^{-1}$ $\leq v \leq 50$~km~sec$^{-1}$ cluster.
While both scenarios lie within the broad range of possibility, 
the first scenario is highly constrained and thus unlikely.    It
is far more likely that the observed nearly symmetrical, low velocity 
distribution includes extraplanar SNRs who have experienced 
moderate to large drag.

Now we turn to \oxyfive\ resonance line emission.   
Observations of the resonance line intensity have the potential
to nicely complement the column density data. The 
intensity produced by a column of \oxyfive\ ions is sensitive
to the local volume density and temperature.
Emissivity goes as density squared.  Hot gas emits readily, while
warm/hot gas emits sparingly and cool gas does not emit.
Using \voyager, \citet{murthyetal} set upper limits on the 
\oxyfive\ doublet intensity for hundreds of sightlines.
Their tightest was a $90\%$ confidence upper limit of 
2600 photons s$^{-1}$ cm$^{-2}$ sr$^{-1}$ toward
$\ell=117^{\rm{o}},b=51^{\rm{o}}$.   Using \fuse,
\citet{shelton01}, \citet{dshl}, \citet{shelton02},
and \citet{otteetal} have detected \oxyfive\ emission from
a handful of high latitude directions outside of 
young SNRs and superbubbles.   Their doublet intensities range from 
$\sim2400$ to $\sim4700$ photons s$^{-1}$ cm$^{-2}$ sr$^{-1}$.
Ignoring the effects of absorption and
subtracting the Local Bubble contribution 
($2\sigma$ upper limit $= 800$ photons s$^{-1}$ cm$^{-2}$ sr$^{-1}$,
\citet{shelton03}) leaves 1600 to 4700 photons s$^{-1}$ cm$^{-2}$ sr$^{-1}$
to be attributed to the extraplanar sky.
In comparison, our model predicts
840 to 2100 photons s$^{-1}$ cm$^{-2}$ sr$^{-1}$, with the
variation mostly due to assumed explosion energy.
Given that the extraplanar SNRs could explain the \oxyfive\
column density between $z = 130$pc and 2000~pc, but not the
higher \oxyfive, the same could be true of the emission.
To determine the answer, we will need to know the typical 
height of the observed \oxyfive\ emission.

\section{Summary}

\begin{itemize}
 \item 
This paper refines Papers I and II's preliminary
calculations of isolated extraplanar SNRs 
by modeling the decrease in interstellar
density with height above the Galactic plane, considering
buoyancy, exploring a wider range of explosion energies
and nonthermal pressures and predicting a wider range
of observables  (\oxyfive, \nitfour, \carthree\ emission
and absorption column densities, $1/4$~keV X-ray surface
brightness, area coverage, volume occupation, vertical
velocity, and variation with height above the Galactic plane).

\item
The predicted average \oxyfive\ column density matches
the observed distribution of column densities as a function of
height above the Galactic plane.   All 9 of our
models (3 sets of explosion energies and
ambient effective magnetic field strengths and 
3 sets of drag coefficients) fell within the scatter in the 
observed $N($\oxyfive$)\sin\left|b\right|$ versus $\left|z\right|$ for
the height range of interest 
($z$ up to 6 times the SN scaleheight).
See Subsection 4.2.

\item
Isolated extraplanar SNRs can explain the
Galactic halo's $1/4$~keV X-ray brightness,
within the precision of the SN rate estimates.
Assuming that these SNRs explode with 
energies of $1 \times 10^{51}$~ergs
and experience an ambient effective magnetic field of $5\ \mu$G,
then the population of isolated SNRs born above the
Galactic \hone\ layer (scaleheight $\sim 130$~pc)
produces a time and space averaged 
$1/4$~keV soft X-ray surface brightness of
$\sim 310 \times 10^{-6}$ \rosat\  counts~s$^{-1}$~arcmin$^{-2}$.
This is $80\%$ of the observationally determined surface brightness
of the Galaxy's southern high latitude sky beyond the
\hone\ layer ($400 \times 10^{-6}$ \rosat\ counts~s$^{-1}$~arcmin$^{-2}$).
If we were to use a larger 
SN explosion rate, such as that estimated by \citet{mw},
then the prediction would rise to 
$400 \times 10^{-6}$ \rosat\ counts~s$^{-1}$~arcmin$^{-2}$,
accounting for the entire observationally determined surface
brightness.   However, using the rate adopted in this paper
and the other choices of SN parameters yields average
surface brightnesses as low as 
$120 \times 10^{-6}$ \rosat\ counts~s$^{-1}$~arcmin$^{-2}$.
Buoyancy makes little difference here.
See Subsection 4.1.


\item 
The predicted average \oxyfive\ 
intensity (840 to 2100 photons s$^{-1}$ cm$^{-2}$ sr$^{-1}$)
contributes significantly to the observationally determined
intensity originating beyond the Local Bubble
($\sim2400$ to $\sim4700$ photons s$^{-1}$ cm$^{-2}$ sr$^{-1}$).
See Subsection 4.2.

\item
Before SNRs develop a cool shell, they are bright
in $1/4$~keV emission, and \oxyfive, \nitfour, and \carthree\
resonance line emission, and are rich in \oxyfive, \nitfour, and \carthree\
ions.   After the shell forms, the remnants dim,
but still harbor large numbers of lithium-like ions.
Thus, the spatial distributions of 
$1/4$~keV emission and \oxyfive\ column densities should be de-correlated,
as found by observers 
(ex: \citet{savageetal03}'s, Figure 9).

 \item
Old SNRs, which glow dimly in $1/4$ keV X-ray and UV resonance line photons and
retain observable numbers of lithium-like ions, 
should cover 
$30$ to $90\%$ of the high latitude sky.
Young remnants, which glow brightly in these regimes
should cover less than a percent of the high latitude sky.
Correspondingly, maps of the observed soft X-ray background 
beyond the Galactic \hone\ layer, outside of superbubbles, 
and after subtracting the extragalactic component appear
mottled, as if roughly half the high latitude sky 
is covered with dim emitting regions while
the other half is bare.   The \oxyfive\ column density surveys
also find null detections and low detections.
See Subsections 3.5, 4.1, and 4.2.

 
\item
The fraction of volume occupied by hot supernova remnant bubbles
peaks around the supernova scaleheight and is less than $10\%$
at all heights.    The volume filled by young remnants is
only a fraction of a percent.
See subsection 3.2.

\item We calculated the vertical displacement and
velocity distribution, assuming
buoyant acceleration and zero to moderate drag.
In the extreme case of unfettered buoyancy,
the velocity can reach $+140$ km sec$^{-1}$, but if SNRs experience
modest drag, then their velocities will range from 0 to 30 km/sec.
In order for the gas to rise and fall again, the drag coefficient
must exceed 1.0.  
Buoyancy is not absolutely necessary to explain the observed
vertical distribution of \oxyfive\ ions.   
See Figure~\ref{fig:buoyancy} and Section 3.3.

\item
Of the explored range of parameters, we prefer
Case 3 ($E_o = 1.0 \times 10^{51}$~ergs, $B_{eff} = 5.0 \mu$G)
because its predictions best match the high ion and soft X-ray
data.   The predictions from the other cases are factors of
$\sim 2$ to $\sim 3$ lower.
Of the explored range of drag coefficients,
we prefer moderate to large drag, because it limits the
buoyant velocity to 50 km sec$^{-1}$ or less.   High latitude
observations find many \oxyfive\ features in this velocity range.

\item This picture can be tested by:
1.) examining the high latitude crescent shaped bright X-ray regions 
($l \sim 247^{\rm{o}}$, $b \sim -64^{\rm{o}}$ 
and $l \sim 215^{\rm{o}}$, $b \sim -68^{\rm{o}}$)
to determine if they are the limbs of young remnants
(see Subsection 4.1),
and 2.) and testing the spectral features of the 
mottled dim $1/4$~keV background for signatures of
old, cooling, recombining SNR gas.
(See Paper II.)

\end{itemize}

\noindent
Acknowledgments:

The author would like to thank Don Cox, K.D. Kuntz, Loris Magnani,
J. Scott Shaw, and Steve Lewis for their contributions to
helpful discussions about the interstellar medium, observations,
and buoyancy.
This work was supported under NASA grant NNG04GD78G.


\clearpage

\begin{thebibliography}{}

\bibitem[Alvarez et al.(97)]{aamo}
	Alvarez, H., Aparici, J., May, J., \& Olmos, F., 1997, A\&AS, 
	124, 315
\bibitem[Avillez \& Mac Low(2001)]{avillezmaclow}
	de Avillez, M. A., \& Mac Low, M.-M.  2001, ApJ, 551, L57
\bibitem[Barber et al.(1996)]{brw}
	Barber, C. R., Roberts, T. P., \& Warwick, R. S., 1996, 
	MNRAS, 282, 157
\bibitem[Benjamin \& Danly(1997)]{benjamin_danly}
	Benjamin, R. A., \& Danly, L. 1997, ApJ, 481, 764
\bibitem[Burrows \& Mendenhall(1991)]{bm}
	Burrows, D. N., \& 
	Mendenhall, J. A., 1991, Science, 351, 629
\bibitem[Caswell \& Lerche(1979)]{cl}
	Caswell, J. L., \& Lerche, I., 1979, MNRAS, 187, 201
\bibitem[Cioffi(1991)]{cioffi}
	Cioffi, D. F. 1991, in International Astronomical Union
	Symposium Proceedings Num. 144, The Disk-halo connection
	In Galaxies, ed. H. Bloemen (the Netherlands), 355
\bibitem[Cox et al.(1999)]{cox-etal-99}
	Cox, D. P., Shelton, R. L., Maciejewski, W., Smith, R. K.,
	Plewa, T. Pawl, A., \& Rozyczka, Michal, 1999, ApJ, 524, 179
\bibitem[Cui \& Cox(1992)]{cui-cox}
	Cui, W. \& Cox, D. P., 1992, ApJ, 401, 206
\bibitem[Cui et al.(1996)]{csmsw}
	Cui, W., Sanders, W. T., McCammon, D., Snowden, S. L., \& 
	Womble, D. S., 1996, \apj, 468, 117
\bibitem[Dickey \& Lockman(1990)]{dl}
	Dickey, J. M., \& Lockman, F. J., 1990, ARA\&A, 28, 215
\bibitem[Dixon, et al.(2001)]{dshl}
	Dixon, W. V., Sallmen, S., Hurwitz, M., \& Lieu, R.
	2001, ApJ, 552, L69
\bibitem[English, et al.(2000)]{englishetal00}
	English, J., Taylor, A. R., Mashchenko, S. Y., Irwin, J. A.,
	Basu, S., \& Johnstone, D. 2000, ApJL, 533, L25
\bibitem[Ferri\`{e}re(1998)]{ferriere98}
	Ferri\`{e}re, K. 1998, ApJ, 503, 700
\bibitem[Ferri\`{e}re(1998)]{ferriere98-2}
	Ferri\`{e}re, K. 1998, ApJ, 503, 759
\bibitem[Ferri\`{e}re \& Zweibel(1991)]{ferriere-zweibel}
	Ferri\`{e}re, K. \& Zweibel, E. G., 1991, ApJ, 383, 602
\bibitem[Garmire et al.(1992)]{gnabfk}
	Garmire, G. P., Nousek, J. A., Apparao, K. M. V., Burrows, D. N., 
	Fink, R. L., \& Kraft, R. P., 1992, 399, 694
\bibitem[Grevesse \& Anders(1989)]{grevesse_anders}
	Grevesse, N., \& Anders, E. 1989, in AIP Conference 
	Proceedings 183,
	Cosmic Abundances of Matter, ed. C. J. Waddington, p. 1
\bibitem[Haslam et al.(1982)]{hssw}
	Haslam, C. G. T., Salter, C. J., Stoffel, H., \& Wilson, W. E.,
	1982, A\&AS, 47, 1
\bibitem[Herbstmeier et al.(1995)]{hmshbmke}
	Herbstmeier, U., Mebold, U., Snowden, S. L., Hartmann, D.,
	Butler Burton, W., Moritz, P., Kalberla, P. M. W., \& Egger, R.,
	1995, A \& A, 298, 606
\bibitem[Howk et al.(2002)]{howketal}
	Howk, J. C., Savage, B. D., Sembach, K. R., \& Hoopes, C. G.
	2002, ApJ, 572, 264
\bibitem[Hurwitz \& Bowyer(1996)]{hurwitzbowyer}
        Hurwitz, M., \& Bowyer, S. 1996, ApJ, 465, 296
\bibitem[Jenkins(1978a)]{jenkins1978a}
	Jenkins, E. B. 1978, ApJ, 219, 845
\bibitem[Jenkins(1978b)]{jenkins1978b}
	Jenkins, E. B. 1978, ApJ, 220, 107
\bibitem[Jenkins \& Meloy(1974)]{jenkinsmeloy}
        Jenkins, E. B., \& Meloy, D. A. 1974, ApJ, 193, 121
\bibitem[Jones(1973)]{jones}
	Jones, E. M. 1973, ApJ, 182, 559
\bibitem[Jones et al.(1996)]{jonesetal}
	Jones, T. W., Ryu, D., \& Tregillis, I. L. 1996, ApJ, 473, 365
\bibitem[Kerp et al.(1993)]{khm}
	Kerp, J., Herbstmeier, U., \& Mebold, U. 1993, A \& A,
	268, L21
\bibitem[Kuntz \& Snowden(2000)]{ks} 
	Kuntz, K. D., \& Snowden, S. L. 2000, \apj, 543, 195
\bibitem[Laming et al.(1996)]{lrmb}
	Laming, J. M., Raymond, J. C., McLaughlin, B. M., \& Blair, W. P.
	1996, \apj 472, 267L
\bibitem[Leahy et al.(1991)]{lnh}
	Leahy, D. A., Nousek, J., \& Hamilton, A. J. S. 1991, \apj 374, 218L
\bibitem[Marshall \& Clark(1984)]{mc}
	Marshall, F. J., \& Clark, G. W., 1984, ApJ, 287, 633
\bibitem[McCammon et al.(1976)]{mmsw}
	McCammon, D., Meyer, S. S., Sanders, W. T., \& Williamson, F. O., 
	1976, ApJ, 209, 46
\bibitem[McCammon et al.(1983)]{mbsk}
	McCammon, D., Burrows, D. N., Sanders, W. T., \& Kraushaar, W. L.,
	1983, ApJ, 269, 107
\bibitem[McKee \& Williams(1997)]{mw}
	McKee, C. F., \& Williams, J. P., 1997, ApJ, 476, 144
\bibitem[Murthy et al.(2001)]{murthyetal}       
	Murthy, J., Henry, R. C., Shelton, R. L., \& Holberg, J. B.
	2001, \apj, 557, 47L
\bibitem[Oegerle et al.(2005)]{oegerleetal}
	Oegerle, W. R., Jenkins, E. B., Shelton, R. L., Bowen, D. V., 
	\& Chayer, P. 2005, ApJ, 622, 377
\bibitem[Otte et al.(2003)]{otteetal}
	Otte, B., Dixon, W. V., \& Sankrit, R. 2003, ApJ, 586, 530
\bibitem[Rosner et al.(1981)]{rabgghmtv}
	Rosner, R., Avni, Y., Bookbinder, J., Giacconi, R., Golub, L., 
	Harnden, F. R., Jr., Maxson, C. W., Topka, K, \& Vaiana, G. S.
	1981, ApJL, 249, L5
\bibitem[Savage et al.(2003)]{savageetal03}
	Savage, B. D., et al., 2003, ApJS, 146, 125
\bibitem[Shelton(1998)]{shelton98}
	Shelton, R. L. 1998, \apj, 504, 785
\bibitem[Shelton(1999)]{shelton99}
	Shelton, R. L. 1999, \apj, 521, 217
\bibitem[Shelton et al.(1999)]{shelton-etal-99}
	Shelton, R. L., Cox, D. P., Maciejewski, W., Smith, R. K.,
	Plewa, T. Pawl, A., \& Rozyczka, Michal, 1999, ApJ, 524, 179
\bibitem[Shelton et al.(2001)]{shelton01}
	Shelton, R. L. et al. 2001, \apj, 560, 730
\bibitem[Shelton (2002)]{shelton02}
	Shelton, R. L. 2002, ApJ, 569, 758
\bibitem[Shelton (2003)]{shelton03}
	Shelton, R. L. 2003, ApJ, 589, 261 
\bibitem[Shelton, Kuntz \& Petre(2004)]{shelton-kuntz-petre}
	Shelton, R. L., Kuntz, K. D., \& Petre, R., 2004, ApJ, 615, 275
\bibitem[Slavin \& Cox(1992)]{slavin-cox-92}
	Slavin, J. D., \& Cox, D. P. 1992, \apj, 392, 131 
\bibitem[Slavin \& Cox(1993)]{slavin-cox-93}
	Slavin, J. D., \& Cox, D. P. 1993, \apj, 417, 187 
\bibitem[Slavin et al.(2000)]{smh}
	Slavin, J. D., McKee, C. F., \& Hollenbach, D. J.
	2000, \apj, 541, 218
\bibitem[Snowden et al.(1991)]{smhhs}
	Snowden, S. L., Mebold, U., 
	Hirth, W., Herbstmeier, U., \& 
	Schmitt, J. H. M. M, 1991, Science, 252, 1529
\bibitem[Snowden et al.(1994)]{smbm}
	Snowden, S. L., McCammon, D., Burrows, D. N., \& Mendenhall, J. A.,
	1994, ApJ, 424, 714
\bibitem[Snowden \& Pietsch(1995)]{sp}
	Snowden, S. L., \& Pietsch, W., 1995, ApJ, 452, 627	
\bibitem[Snowden et al.(1997)]{sefmpsstv}
	Snowden, S. L., Egger, R., Freyberg, M. J., McCammon, D., 
	Plucinsky, P. P., 1998, 
	Sanders, W. T., Schmitt, J. H. M. M., Tr{\"{u}}mper, \& J., Voges, 
	1997, ApJ, 485, 125
\bibitem[Snowden et al.(1998)]{seffp}
	Snowden, S. L., Egger, R., 
	Finkbeiner, D., Freyberg, M. J., \& Plucinsky, P. P., 
	1998, ApJ, 493, 715
\bibitem[Snowden et al.(2000)]{sfks}
	Snowden, S. L., Freyberg, M. J., Kuntz, K. D., \& Sanders, W. T.,
	2000, ApJ, ApJS, 128, 171
\bibitem[Wakker et al.(2003)]{wakkeretal03}
	Wakker, B. P. et al., 2003, ApJS, 146, 1
\bibitem[Wang \& Ye(1996)]{wy}
	Wang, Q. D., \& Ye, T., 1996, New Astronomy, 1, 245
\bibitem[Widmann et al.(1998)]{widmannetal}
	Widmann, H., et al. 1998, A \& A, 338, L1
\bibitem[Wolfire et al.(1995)]{wolfire_etal}
	Wolfire, M. G., McKee, C. F., Hollenbach, D., \& Tielens, A. G. G. M,
	ApJ, 453, 673
\bibitem[Zsarg\'o et al.(2003)]{zsargoetal}
	Zsarg\'o, J., Sembach, K. R., Howk, J. C., \& Savage, B. D.
	2003, ApJ, 586, 1019

 \end{thebibliography}
\end{document}